\documentclass[sts]{imsart}

\usepackage{subcaption}
\usepackage{threeparttable}
\DeclareCaptionSubType*[Alph]{table}
\DeclareCaptionLabelFormat{mystyle}{Table~\bothIfFirst{#1}{ ̃}#2}
\RequirePackage[OT1]{fontenc}
\RequirePackage{amsthm,amsmath}

\RequirePackage[colorlinks,citecolor=blue,urlcolor=blue]{hyperref}

\usepackage{graphicx,psfrag,epsf}
\usepackage{enumerate}

\usepackage{url} 

\arxiv{arXiv:1210.0870}

\startlocaldefs
\numberwithin{equation}{section}
\theoremstyle{plain}

\endlocaldefs

\begin{document}

\begin{frontmatter}

\title{Maximum likelihood multiple imputation: Faster imputations and consistent standard errors without posterior draws}
\runtitle{Maximum likelihood multiple imputation}

\begin{aug}
  \author{\fnms{Paul T.}  \snm{von Hippel}\ead[label=e1]{paulvonhippel.utaustin@gmail.com}}
  \and \author{\fnms{Jonathan W.} \snm{Bartlett}\ead[label=e2]{j.w.bartlett@bath.ac.uk}}

  \runauthor{von Hippel \& Bartlett}

  \affiliation{University of Texas, Austin, USA and University of Bath, UK}

\end{aug}

\begin{abstract}
Multiple imputation (MI) is a method for repairing and analyzing data with missing values. MI replaces missing values with a sample of random values drawn from an \textit{imputation model}. 
The most popular form of MI, which we call \textit{posterior draw multiple imputation} (PDMI), draws the parameters of the imputation model from a Bayesian posterior distribution. An alternative, which we call \textit{maximum likelihood multiple imputation} (MLMI), estimates the parameters of the imputation model using maximum likelihood (or equivalent). Compared to PDMI, MLMI is less computationally intensive, faster, and yields slightly more efficient point estimates.

A past barrier to using MLMI was the difficulty of estimating the standard errors of MLMI point estimates. We derive, implement, and evaluate three consistent standard error formulas: (1) one combines variances within and between the imputed datasets, (2) one uses the score function, and (3) one uses the bootstrap to estimate variance components due to sampling and imputation. Formula (1) modifies for MLMI a formula that has long been used under PDMI, while formulas (2) and (3) can be used without modification under either PDMI or MLMI. We have implemented MLMI and the standard error estimators in the \textit{mlmi} and \textit{bootImpute} packages for R. 
\end{abstract}


\end{frontmatter}


\section{Introduction}

\textit{Multiple imputation} (MI) is a popular method for repairing and analyzing data with missing values \cite{Rubin1987}. Under MI, the distribution of missing values is assumed to depend on the observed values $Y_{obs}$ and an imputation model with parameter vector $\theta $. Then MI proceeds in two steps:

\begin{enumerate}
\item  Obtain a parameter estimate ${\widehat{\theta }}_{obs,m}$ from $Y_{obs}$ alone.

\item  Fill in each missing value with a random imputation drawn conditionally on $Y_{obs}$ and ${\widehat{\theta }}_{obs,m}$.
\end{enumerate}

\noindent These steps iterate multiple times ($m$=1,{\dots},$M$), returning $M$ imputed copies of the dataset. These MI data are analyzed to produce an MI point estimate ${\widehat{\theta }}_{MI}$ and an estimate of its variance $V_{MI}=V(\widehat{\theta }_{MI})$. Under some circumstances, discussed later \cite{Bartlett2015}, MI data can also be analyzed to estimate additional quantities that are not the same as the imputation parameters $\theta$.

Different estimates ${\widehat{\theta }}_{obs,m}$ can be used for $\theta$. The most common approach draws ${\widehat{\theta }}_{obs,m}$ at random from the Bayesian posterior distribution of the parameters given $Y_{obs}$ \cite{Rubin1987}. We call estimates drawn in this way \textit{posterior draws} (PD), or ${\widehat{\theta }}_{PD,m}$, and when PD estimates are used in the imputation model, we call the approach \textit{posterior draw multiple imputation}.

An alternative is to estimate the imputation parameters by applying \textit{maximum likelihood} (ML) to the incomplete data $Y_{obs}$ \cite{Wang1998,Robins2000,Kim2009,vH2013,vH2015}. Imputation parameters estimated in this way are ML estimates, ${\widehat{\theta }}_{ML}$, and when ML estimates are used in the imputation model, we call the approach \textit{maximum likelihood multiple imputation} (MLMI). Any approach that uses asymptotically efficient estimates of the imputation parameters is equivalent to MLMI.

Although PDMI is by far the more common approach in practice, it does have certain disadvantages. A minor disadvantage is that PDMI point estimates are less efficient than MLMI point estimates, but the difference in efficiency is trivial unless the fraction of missing information is large and the number of imputations $M$ is very small \cite{Wang1998}. Likewise, point estimates can have more small-sample bias under PDMI than under MLMI, but the biases are trivial in moderate to large samples \cite{vH2013,vH2015}.

The more serious disadvantage of PDMI is computational. PDMI software users sometimes report runtimes or hours or even days in large datasets \cite{Eid2016, Rojas2012, Huang2015, Lanning2003}. Although increases in computing power should have speeded PDMI up, in practice these increases have been offset by growth in the size of datasets and growth in the recommended number of imputations $M$. In early MI research, $M=3-10$ imputations were recommended as adequate for stable point estimates \cite{Rubin1987}, but more recent research, evaluating the stability of standard error estimates and confidence intervals, calls for as many as $M=20-200$ imputations in data with a high fraction of missing information \cite{vH2018}.

In addition to long runtimes, PDMI software can be "fussy," sometimes failing to converge \cite{Honaker2010}, giving errors and warning messages that seem inscrutable to end users \cite{Rojas2012}, or requiring diagnostics and changes to the prior distribution that few end users, or even experts, are qualified to carry out \cite{Schafer1997, SAS2000, Honaker2010}. Long runtimes and convergence issues contribute to the impression---not uncommon among applied researchers---that MI is not worth the trouble. This limits the adoption of MI, which is still rare in some applied fields, such as economics.

Most of these problems occur because most PDMI software uses a computationally intensive Markov Chain Monte Carlo (MCMC) algorithm known as \textit{data augmentation} \cite{Schafer1997}. Faster algorithms are available to get PD estimates \cite{King2001}, and estimation can be further accelerated by running the algorithm in parallel \cite{SSCC2012, Honaker2015}. But efforts to speed up PD estimates beg the question of whether we need PD estimates at all. 

Can't we do imputation without posterior draws, as MLMI does? No matter what we do to speed up PDMI, MLMI will always be faster, and MLMI point estimates will always be more efficient. Why, then, hasn't MLMI been used more often? 

A major barrier to MLMI's adoption has been a lack of convenient formulas for estimating the variance of MLMI point estimates. The variance of PDMI point estimates can be estimated by a simple \textit{within-between} (WB) formula \eqref{V_PDMI_WB} that combines variances within and between the imputed datasets \cite{Rubin1987}. But that WB formula, when applied to MLMI data, will produce variance estimates that are too small on average. For that reason, MLMI has been labeled ``improper'' \cite{Rubin1987}, and perhaps that label has discouraged investigation. Alternative formulas have been proposed for variance estimation under MLMI \cite{Wang1998,Robins2000}, but the formulas are cumbersome and require statistical quantities that are often unavailable in applied data analysis.

In this article, we make MLMI more usable by deriving three simpler estimators for the variance of MLMI point estimates. One formula \eqref{V_MLMI_WB} modifies the WB formula that is used with PDMI. One formula \eqref{V_MI|SB} simplifies a score-based (SB) variance formula first proposed by Wang \& Robins \cite{Wang1998}. And one formula \eqref{Vhat_BMI} combines MI with the bootstrap to calculate variance components due to sampling and imputation. We have implemented these estimators in the \textit{mlmi} and \textit{bootImpute} packages for R, which we have published on the Comprehensive R Archive Network (CRAN) \cite{Bartlett2019a,Bartlett2019b}.

With these new variance formulas, MLMI becomes a more practical alternative to PDMI. The rest of this article derives the variance estimators, compares their properties analytically and through simulation, and demonstrates their use in an applied data analysis. 

\section{Incomplete data}

Before describing different estimators, let's define the missing data problem. 

If we had complete data $Y_{com}$ with $N$ cases, we could maximize its likelihood to get a complete-data ML estimate ${\widehat{\theta }}_{com}$ of the parameter vector $\theta$. But instead we have incomplete data where some values $Y_{mis}$ are missing and other values $Y_{obs}$ are observed. If values are missing at random (MAR)---so that the probability of a value being missing depends only on $Y_{obs}$---then we can get a consistent ML estimate ${\widehat{\theta }}_{ML}$ using only $Y_{obs}$, without  modeling the process that causes values to be missing \cite{Rubin1976}. Note that ${\widehat{\theta }}_{ML}$ is calculated from \textit{all} the observed values, including observed values in cases with missing values \cite{Dempster1977, Arbuckle1996}.

The variance $V_{ML}=V({\widehat{\theta }}_{ML})$ of the observed-data ML estimate exceeds the variance $V_{com}=V({\widehat{\theta }}_{com})$ that we would get if we had complete data. So the information $V^{-1}_{ML}$ in in the observed data is less than the information $V^{-1}_{com}$ that the complete data would provide. The difference is the \textit{missing information}:
\begin{equation}
V^{-1}_{mis}=V^{-1}_{com} - V^{-1}_{ML}
\end{equation}
The ratio of observed to complete information is the \textit{fraction of observed information} ${\gamma }_{obs}$, and the ratio of missing to complete information is the \textit{fraction of missing information} ${\gamma }_{mis}$:
\begin{eqnarray}
{\gamma }_{obs} &=& V^{-1}_{ML}V_{com} \\
{\gamma }_{mis} &=& V^{-1}_{mis}V_{com} = I-{\gamma }_{obs}
\end{eqnarray}

If $\theta$ is a scalar, then these variances and fractions are scalars. If $\theta$ is a vector, then these ``variances'' are covariance matrices, and the fractions of observed and missing information are matrices as well.

\section{Multiple imputation}

MI is an algorithm with $M$ iterations. In iteration $m=1,\dots ,M$, MI carries out the following steps:

\begin{enumerate}
\item  From the observed data $Y_{obs}$, obtain an observed-data estimate ${\widehat{\theta }}_{obs,m}$. 

\item  Fill in the missing data $Y_{mis}$ with random imputations $Y_{imp,m}$ drawn conditionally on $Y_{obs}$ and ${\widehat{\theta }}_{obs,m}$. The result is a singly imputed (SI) data set $Y_{SI,m}=\{Y_{obs},Y_{imp,m}\}$.
\end{enumerate}

\noindent Together, the $M$ SI datasets make up an MI dataset $Y_{MI}$.

The difference between MLMI and PDMI lies in the definition of the observed-data estimator ${\widehat{\theta }}_{obs,m}$ in step 1: 
\begin{itemize}
\item Under MLMI, ${\widehat{\theta }}_{obs,m}$ is the ML estimate ${\widehat{\theta }}_{ML}$, or another estimate that just as efficient in large samples.

\item Under PDMI, ${\widehat{\theta }}_{obs,m}$ is a PD estimate ${\widehat{\theta }}_{PD,m}$ drawn at random from the posterior distribution of $\theta $ given $Y_{obs}$.
\end{itemize}

\subsection{ Computational efficiency of MLMI over PDMI}
\label{section:computational_efficiency}

The main advantage of MLMI is its computational efficiency. Under PDMI, a new PD estimate ${\widehat{\theta }}_{PD,m}$ must be drawn in every iteration $m$, so both steps of the algorithm must be iterated. Under MLMI, by contrast, the observed-data ML estimate ${\widehat{\theta }}_{ML}$ is the same in every iteration, so we can run step 1 just once and only iterate step 2. Not iterating step 1 gives MLMI a speed advantage that increases with the number of iterations $M$.  

Even when $M$ is small, MLMI remains faster because it is faster to get ML estimates than it is to get PD estimates. In some simple settings (such as our simulation, later), both ML and PD estimates can be calculated using closed-form formulas; PD requires an extra step, but the extra runtime is trivial. In general settings, though, both ML and PD estimates require iterative, numerical methods, which are much more computationally intensive for PD than for ML. To get ML estimates, software can use \textit{full information maximum likelihood} or the EM algorithm \cite{Enders2001}. But to get PD estimates, most PDMI software uses data augmentation \cite{Schafer1997}, in which the EM algorithm is only the first step. Data augmentation typically begins by using the EM algorithm to find the posterior mode of the parameters of the imputation model. It then takes a random walk around the posterior by iteratively re-imputing the data and re-estimating the imputation parameters from the imputed data. The re-estimated parameters are PD estimates.

The main reason why data augmentation is slow to return results is that it discards results from the vast majority of iterations. It discards (``burns in''), say, the first 100 iterations to ensure that the PD estimates have converged to their posterior distribution; then it discards, say, 99 out out every 100 PD estimates, to ensure that the PD estimates are approximately uncorrelated. So 100$M$ iterations may be required to get $M$ PD estimates and $M$ imputed datasets.

A faster and stabler way to get PD estimates is to bootstrap the incomplete data and calculate an ML estimate from each bootstrapped sample \cite{vanBuuren2018a, Heitjan1991, King2001}. Both data augmentation and bootstrapped ML are faster if they run in parallel \cite{SSCC2012, Honaker2015}. But both remain slower than ML, and PDMI remains slower than MLMI.

\subsection{Bootstrapped MI}
A variant of MI which can be useful for variance estimation is \textit{bootstrapped MI} (BMI). BMI is an iterative procedure with two nested loops. In iteration $b=1,\dots,B$, 

\begin{enumerate}
\item Take a bootstrapped sample $Y_{boot,b}$ of $N$ cases from the incomplete data.
\item Then, in iteration $d=1,\dots,D$, apply MI to $Y_{boot,b}$. That is,
\begin{enumerate}
\item  From the observed values in $Y_{boot,b}$, obtain an observed-data estimate ${\widehat{\theta }}_{obs,bd}$. 

\item  Fill in $Y_{boot,b}$'s missing values with random imputations drawn conditionally on ${\widehat{\theta }}_{obs,bd}$ and the observed values in $Y_{boot,b}$. The result is a single bootstrapped-then-imputed (BSI) dataset $Y_{BSI,bd}$.
\end{enumerate}
\end{enumerate}

\noindent Together, the $BD$ BSI datasets make up an BMI dataset $Y_{BMI}$.

There are two flavors of BMI: boostrapped MLMI (BMLMI) and bootstrapped PDMI (BPDMI). The difference is the definition of the estimator ${\widehat{\theta }}_{obs,bd}$: 
\begin{itemize}
\item Under BMLMI, ${\widehat{\theta }}_{obs,bd}$ is an ML estimate ${\widehat{\theta }}_{ML,b}$ derived from the observed values in $Y_{boot,b}$.

\item Under BPDMI, ${\widehat{\theta }}_{obs,bd}$ is a PD estimate ${\widehat{\theta }}_{PD,bd}$ drawn at random from the posterior distribution of $\theta $ given the observed values in $Y_{boot,b}$.
\end{itemize}

As in other applications of the bootstrap, $B=40$ is adequate for some purposes, though larger $B$ is better. The optimal value for $D$, however, is 2, for reasons we will discuss when we get to variance estimation.

Just as MLMI is faster than PDMI, BMLMI is faster than BPDMI. Not only is ${\widehat{\theta }}_{ML,b}$ easier to calculate than ${\widehat{\theta }}_{PD,bd}$, but ${\widehat{\theta }}_{ML,b}$ only needs to be calculated once for each bootstrapped sample, while ${\widehat{\theta }}_{PD,bd}$ needs to be calculated $M$ times for each bootstrapped sample. That is, in the $b^{th}$ bootstrapped sample, PDMI must iterate all of step 2, while MLMI can run step 2(a) just once and only iterate step 2(b).

\section{MI point estimates}

With large $N$, $M$, and $BD$, practically equivalent point estimates can be calculated from data that was imputed using PDMI or MLMI, with or without the bootstrap. With modest $M$ or $BD$, however, MLMI point estimates are more efficient than PDMI point estimates, and point estimates from either MLMI or PDMI are more efficient without the bootstrap than with it. This section shows why. 

There are several ways to get point estimates from MI data. The most common way is \textit{repeated MI} \cite{Rubin1987}, which analyzes each SI dataset as though it were complete, producing $M$ SI point estimates ${\widehat{\theta }}_{SI,m},m=1,\dots ,M$, whose average is a repeated MI point estimate:
\begin{equation}
\widehat{\theta }_{MI}=\frac1{M}\sum_{m=1}^{M} \widehat{\theta }_{SI,m}
\label{thetaHatMI}
\end{equation}

\noindent Under MLMI we call this estimate ${\widehat{\theta }}_{MLMI}$; under PDMI we call it ${\widehat{\theta }}_{PDMI}$. The corresponding SI estimators are ${\widehat{\theta }}_{MLSI}$ and ${\widehat{\theta }}_{PDSI}$. The limit of $\widehat{\theta }_{MI}$ as $M$ gets large is $\lim_{M\to\infty}{\widehat{\theta }_{MI}={\theta }_{\infty I}}$.

A rarely used alternative is \textit{stacked MI}, in which the MI datasets are stacked and analyzed as though they represented a single dataset with $NM$ observations. In large samples, stacked MI and repeated MI yield equivalent point estimates \cite{Wang1998}, but repeated MI is more convenient for variance estimation.

We can also get point estimates from BMI data. Analyze each of the bootstrapped-then-imputed datasets as though it were complete to obtain \textit{BD} individual point estimates ${\widehat{\theta }}_{BD}$. Then average the individual estimates $\widehat{\theta }_{BD}$ to get a BMI point estimate:
\begin{equation} 
\widehat{\theta }_{BMI}=\frac1{BD}\sum^B_{b=1}{\sum^D_{d=1}\widehat{\theta }_{bd}} 
\end{equation} 

\noindent Under BMLMI we call this estimate ${\widehat{\theta }}_{BMLMI}$; under BPDMI we call it ${\widehat{\theta }}_{BPDMI}$.

\subsection{Variance of MI point estimates}
\label{section:V_MI}

Repeated MI point estimates ${\widehat{\theta }}_{MI}$ are consistent, asymptotically normal, and approach ${\widehat{\theta }}_{ML}$ as $M$ and $N$ get large. This is true under both MLMI and PDMI. With large $N$, the variance of an MI point estimate is \cite{Wang1998}
\begin{equation}
\label{V_MI}
V_{MI}=V_{ML}+\frac1{M}(V_{SI}-V_{ML})
\end{equation}

Although MLMI and PDMI point estimates have approximately the same variance when $M$ is large, when $M$ is finite, MLMI point estimates have smaller variance than PDMI point estimates. To understand why, notice that the variance of an MI point estimate depends to some degree on the variance of the underlying observed-data estimate $\widehat{\theta }_{obs,m}$---and in large samples ML estimates have the smallest variance possible. In fact, PD estimates are approximately twice as variable as ML estimates \cite{vH2013, vH2015}. To see this, notice that $\widehat{\theta }_{PD,m}$ is drawn from a posterior density whose asymptotic distribution is $\widehat{\theta }_{PD,m}\sim N(\widehat{\theta }_{ML},\widehat{V}_{ML}$). So the variance of $\widehat{\theta }_{PD,m}$ is $V_{PD}=V(\widehat{\theta }_{ML})+\widehat{V}_{ML}\approx 2V_{ML}$.

The substantial efficiency advantage of $\widehat{\theta }_{ML}$ over $\widehat{\theta }_{PD}$ translates into a smaller efficiency advantage of $\widehat{\theta }_{MLMI}$ over $\widehat{\theta }_{PDMI}$. With large $N$, the variances of $\widehat{\theta }_{MLMI}$ and $\widehat{\theta }_{PDMI}$ are
\begin{equation}
\label{V_MLMI}
V_{MLMI}=V({\widehat{\theta }}_{MLMI}) = V_{ML}+\frac1{M}V_{com}{\gamma }_{mis}
\end{equation}
\begin{equation}
V_{PDMI}=V({\widehat{\theta }}_{PDMI})  = V_{ML}+\frac1{M}V_{ML}{\gamma }_{mis}
\label{V_PDMI}
\end{equation}
These expressions come from Wang and Robins (1998, equations 1 and 2), but we have simplified the expression for $V_{PDMI}$; the steps of the simplification are given in Appendix \ref{appendix: simplify_V_PDMI}. 

Since $V_{com}<V_{ML}$ it follows that $V_{MLMI}<V_{PDMI}$---that is, MLMI is more efficient than PDMI in large samples. In small samples, MLMI is also more efficient and less biased than PDMI, at least in normal data \cite{vH2013,vH2015}.

Later it will be helpful to have expressions for the variance of the SI estimators. We can get those expressions by taking the variance of the MI estimators and setting $M$ = 1:
\begin{equation}
V_{MLSI}=V({\widehat{\theta }}_{MLSI}) \xrightarrow{N \to \infty} V_{ML}+V_{com}{\gamma }_{mis}
\label{V_MLSI} 
\end{equation}
\begin{equation}
V_{PDSI}=V({\widehat{\theta }}_{PDSI})  \xrightarrow{N \to \infty} V_{ML}+V_{ML}{\gamma }_{mis}
\label{V_PDSI}
\end{equation}

\subsection{Variance of BMI point estimates}
The variance of BMI point estimates is a little different. It can be calculated as follows. In large samples, the individual bootstrapped-then-imputed point estimates ${\widehat{\theta }}_{bd}$ fit a random effects model that is centered around ${\widehat{\theta }}_{ML}$:
\begin{equation}
\label{bootstrap_re_model}
\widehat{\theta }_{bd}=\widehat{\theta }_{ML} +e_b+e_{bd} 
\end{equation}

\noindent where $e_b$ represents bootstrap or sampling variation, and $e_{bd}$ represents imputation variation. The variance components are 
\begin{equation}
V(e_b)=V_{ML} 
\end{equation}

\noindent and
\begin{eqnarray}
\label{V_e_bd}
V(e_{bd})=V_{SI}-V_{ML} 
&=& \begin{cases}
   V_{com} \gamma_{mis} & \text{under BMLMI} \\    
   V_{ML} \gamma_{mis} & \text{under BPDMI} 
\end{cases}
\end{eqnarray}

\noindent where the expressions in the final brace come from substituting \eqref{V_MLSI} and \eqref{V_PDSI} for $V_{SI}$. 

The BMI point estimate is just the average $\frac1{BD}\sum\sum\widehat{\theta }_{bd}$, so its variance is
\begin{equation}
\label{V_BMI}
V_{BMI} = V(\widehat{\theta }_{BMI})=V_{ML}+\frac{V_{ML}}{B}+\frac{V_{SI}-V_{ML}}{BD} 
\end{equation}

\noindent Clearly $V_{BMI}$ decreases faster with $B$ than with $D$, so it makes sense to set $D$ as low as possible. We recommend $D=2$ since at least 2 imputations per bootstrap sample are needed for variance estimation.

With $B$ bootstrap samples each imputed $D$ times, a $\widehat{\theta }_{BMI}$ point estimate is more variable than a non-bootstrapped MI point estimate $\widehat{\theta }_{MI}$ with $M=BD$ imputations. The difference in variance
\begin{equation}
\label{V_BMI_vs_V_MI}
V_{BMI}-V_{MI}=\frac{V_{ML}}{B} 
\end{equation}
\noindent is obtained by subtracting \eqref{V_MI} from \eqref{V_BMI} with $BD=M$. Again, it is clear that $V_{BMI}$ is smaller when $B$ is large and $D=M/B$, perforce, is small. That is one reason we recommend setting $D=2$.

The variance of a BMI point estimates is smaller under BMLMI than under BPDMI. We get the following expressions by substituting \eqref{V_MLSI} and \eqref{V_PDSI} for $V_{SI}$ in \eqref{V_BMI}:
\begin{equation}
\label{V_BMLMI}
V_{BMLMI} = 
   V_{ML}\left(1+\frac1{B}\right)+\frac1{BD}V_{com} \gamma_{mis} \text{under BMLMI} 
\end{equation}
\begin{equation}
\label{V_BPDMI}
V_{BPDMI} = 
   V_{ML}\left(1+\frac1{B}\right)+\frac1{BD} V_{ML} \gamma_{mis}  \text{under BPDMI} 
\end{equation}

\noindent Since $V_{com}<V_{ML}$, it follows that $V_{BMLMI} < V_{BPDMI}$.

\subsection{ How many imputations are needed for point estimates?}

How many imputations are needed to produce MI point estimates that are almost as efficient as they would be with infinite imputations? The answer depends on the fraction of missing information ${\gamma }_{mis}$ and on whether MLMI or PDMI is used. The large-$N$ efficiencies of ${\widehat{\theta }}_{MLMI}$ and ${\widehat{\theta }}_{PDMI}$, relative to ${\widehat{\theta }}_{ML}$, are
\begin{eqnarray}
re_{MLMI}=V^{-1}_{MLMI}V_{ML} &=& \left(I+\frac1{M}{\gamma }_{obs}{\gamma }_{mis}\right)^{-1} \\
re_{PDMI}=V^{-1}_{PDMI}V_{ML} &=& \left(I+\frac1{M}{\gamma }_{mis}\right)^{-1}
\end{eqnarray}

\noindent These relative efficiencies were calculated from expressions \eqref{V_MLMI} and \eqref{V_PDMI}. The expression for $re_{PDMI}$, derived a different way, also appears in Rubin \cite{Rubin1987}, p. 114.\footnote{Rubin was estimating the efficiency of a PDMI point estimate with $M$ imputations relative to one with infinite imputations, whereas we are calculating the efficiency of a PDMI estimate relative to an ML estimate. In large samples, however, an ML estimate is equivalent to a PDMI estimate with infinite imputations, so the two definitions of asymptotic efficiency are the same.}

Under BMI, the efficiencies of ${\widehat{\theta }}_{BMLMI}$ and ${\widehat{\theta }}_{BPDMI}$, relative to ${\widehat{\theta }}_{ML}$, are
\begin{eqnarray}
re_{BMLMI}=V^{-1}_{MLMI}V_{ML} &=& \left(\left(1+\frac1{B}\right)I+\frac1{BD}{\gamma }_{obs}{\gamma }_{mis}\right)^{-1} \\
re_{BPDMI}=V^{-1}_{PDMI}V_{ML} &=& \left(\left(1+\frac1{B}\right)I+\frac1{BD}{\gamma }_{mis}\right)^{-1}
\end{eqnarray}
\noindent These efficiencies were calculated from \eqref{V_BMI}.

Table~\ref{tab:point_estimates_imputations} shows the number of imputations that are needed for MI point estimates to have 95\% asymptotic relative efficiency. Under MI the number of imputations is $M$; under BMI, it is $BD$ with $D=2$. 

\begin{table}
\caption{Number of imputations needed for point estimates to have 95\% asymptotic relative efficiency.}
\label{tab:point_estimates_imputations}
\begin{tabular}{ c c c c c } \hline 
 & \multicolumn{4}{c}{Imputations needed} \\ \cline{2-5}
${\gamma }_{mis}$ & PDMI & MLMI & BPDMI & BMLMI \\ \hline
\cline{2-5}
.1&2&2&38&36 \\
.2&4&3&38&38 \\
.3&6&4&40&38 \\
.4&8&4&42&40 \\
.5&10&4&44&40 \\
.6&12&4&46&40 \\
.7&14&4&48&38 \\
.8&16&3&50&38 \\
.9&18&2&50&36 \\
 \hline
\end{tabular}

\textit{Note}. For PDMI and MLMI, the number of imputations shown is $M$. For BPDMI \& BMLMI, the number of imputations shown is $BD$, where $B$ is the number of bootstrap samples and $D=2$ is the number of imputations per bootstrap sample.
\end{table}

MLMI point estimates need fewer imputations than PDMI point estimates, especially when ${\gamma }_{mis}$ is large. Under PDMI, the number of imputations needed increases linearly as $M=2{\gamma }_{mis}$, but under MLMI, $M$ is a quadratic function of ${\gamma }_{mis}$ that peaks at $M=4$ near ${\gamma }_{mis}=.5$ and falls if ${\gamma }_{mis}$ is larger or smaller. PDMI and MLMI need similar numbers of imputations if ${\gamma }_{mis}$ is small, but if ${\gamma }_{mis}$ is large MLMI needs many fewer imputations. For example,if ${\gamma }_{mis}=.9$, MLMI needs just 2 imputations while PDMI needs 18 imputations to achieve the same efficiency.

Under BMI, BMLMI needs fewer imputations than BPDMI to achieve point estimates with the same efficiency. But the difference is relatively small. Using either form of BMI, 38 to 50 imputations typically suffice---i.e., 19 to 25 bootstrapped datasets, each imputed twice.

If the efficiency of point estimates were all that mattered, we would clearly choose MLMI over PDMI, and we wouldn't give BMI a second thought. But the picture changes somewhat when we go beyond point estimates and consider variance estimates as well.

\section{WB variance estimates}
\label{section:WB}

In the coming sections, we will derive three ways to estimate the variance of an MI point estimate. We call these the WB variance estimate, the SB variance estimate, and the bootstrapped MI variance estimate. Each variance estimate can be used to calculate a confidence interval (or hypothesis test) and estimate the fraction of missing information. Both WB and SB estimates make certain assumptions about the imputation and analysis model, which we'll discuss later. bootstrapped MI makes fewer assumptions.

This section derives the \textit{within-between} (\textit{WB) estimators}, so called because they rely on variance components that lie within and between the SI datasets in MI data.

When we analyze an SI dataset as though it were complete, we get not just an SI point estimate ${\widehat{\theta }}_{SI,m}$ but also an SI variance estimate ${\widehat{V}}_{com,SI,m}$ that would consistently estimate the variance if the data were complete. Across the $M$ SI datasets, the average of the ${\widehat{V}}_{com,SI,m}$ is the within variance ${\widehat{W}}_{MI}$, and the variance of the SI point estimates ${\widehat{\theta }}_{SI,m}$ is the between variance ${\widehat{B}}_{MI}$.
\begin{eqnarray}
{\widehat{W}}_{MI} &=& \frac1{M}\sum^M_{m=1}{{\widehat{V}}_{com,SI,m}} \\
\widehat{B}_{MI} &=& \frac1{M-1}\sum^M_{m=1}({\widehat{\theta }_{SI,m}-\widehat{\theta }_{MI}})^{\otimes{2}}
\end{eqnarray}

\noindent Here the notation $({\widehat{\theta }}_{SI,m}-\widehat{\theta }_{MI})^{\otimes 2}$ represents the outer product $(\widehat{\theta }_{SI,m}-\widehat{\theta }_{MI})(\widehat{\theta }_{SI,m}-\widehat{\theta }_{MI})^T$, which reduces to the square $(\widehat{\theta }_{SI,m}-\widehat{\theta }_{MI})^2$ if $\theta$ is scalar \cite{Wang1998}.

Clearly $\widehat{W}_{MI}$ is a consistent estimator of $V_{com}$ \cite{Rubin1987, Tsiatis2006}. ${\widehat{B}}_{MI}$ is an unbiased and consistent estimator for the variance of ${\widehat{\theta }}_{SI}$ around ${\widehat{\theta }}_{\mathrm{\infty }I}$, and since ${\widehat{\theta }}_{\mathrm{\infty }I}$ approaches ${\widehat{\theta }}_{ML}$ in large samples, it follows that ${\widehat{B}}_{MI}$ consistently estimates

\begin{eqnarray}
\label{EB_MI}
E(\widehat{B}_{MI}) &=& V(\widehat{\theta }_{SI} \mid \widehat{\theta }_{{\infty}I})  \xrightarrow{N \to \infty} V(\widehat{\theta }_{SI} \mid \widehat{\theta }_{ML})  \\
&=& V_{SI} - V_{ML} \nonumber \\
&=& \begin{cases}
    V_{com} \gamma_{mis} & \text{under MLMI} \nonumber \\    
    V_{ML} \gamma_{mis}  & \text{under PDMI} 
\end{cases}
\end{eqnarray}

The last line, which is obtained by substituting expressions \eqref{V_MLSI} and \eqref{V_PDSI} for $V_{SI}$, shows that ${\widehat{B}}_{MI}$ estimates a different quantity under MLMI than under PDMI. When this distinction is important, we will use the symbols ${\widehat{B}}_{MLMI}$ and ${\widehat{B}}_{PDMI}$, along with ${\widehat{W}}_{MLMI}$ and ${\widehat{W}}_{PDMI}$.

A useful corollary of \eqref{EB_MI} is that ${\widehat{B}}_{MI}/M$ is a consistent estimator for the variance of ${\widehat{\theta }}_{MI}$ around ${\widehat{\theta }}_{ML}$:

\begin{eqnarray}
E\left(\frac1{M}\widehat{B}_{MI}\right) \xrightarrow{N \to \infty} \frac1{M} V(\widehat{\theta }_{SI} \mid \widehat{\theta }_{ML}) &=& V(\widehat{\theta }_{MI} \mid \widehat{\theta }_{ML})  \\
&=& V_{MI} - V_{ML} \nonumber
\label{EB_MI_over_M}
\end{eqnarray}

\noindent So if we derive a consistent estimator of $V_{ML}$, we can add ${\widehat{B}}_{MI}/M$ to get a consistent estimator of $V_{MI}$. 

Although consistent, ${\widehat{B}}_{MI}$ can be imprecise when $M$ is small, because ${\widehat{B}}_{MI}$ is a variance estimated from a sample of just $M$ imputations. Estimators that give substantial weight to ${\widehat{B}}_{MI}$ will be imprecise as well. We will return to this issue repeatedly in the next couple of pages.

\subsection{ Under PDMI}

Under PDMI, the WB variance estimator is
\begin{equation}
{\widehat{V}}_{PDMI,WB}={{\widehat{W}}_{PDMI}+\widehat{B}}_{PDMI}+\frac1{M}{\widehat{B}}_{PDMI}
\label{V_PDMI_WB}
\end{equation}

\noindent This estimator can be derived in a Bayesian framework \cite{Rubin1987}, but it can also be derived by substituting consistent estimators for the components of $V_{PDMI}$ in equation \eqref{V_PDMI} \cite{Wang1998}. That is, ${\widehat{V}}_{PDMI,WB}$ consistently estimates $V_{PDMI}$ because ${\widehat{W}}_{PDMI}$ consistently estimates $V_{com}$, ${\widehat{B}}_{PDMI}$ consistently estimates $V_{ML}-V_{com}$, and ${\widehat{B}}_{PDMI}/M$ consistently estimates ${V_{PDMI}-V}_{ML}$.

These are WB estimators for the fractions of observed and missing information under PDMI:
\begin{eqnarray}
\widehat{\gamma }_{obs|PDMI,WB} &=& (\widehat{W}_{PDMI}+\widehat{B}_{PDMI})^{-1}\widehat{W}_{PDMI} \\
\widehat{\gamma }_{mis|PDMI,WB} &=& I-\widehat{\gamma }_{obs|PDMI,WB}
\end{eqnarray}

\noindent Again the consistency of these estimators can be verified by substitution. ${\widehat{\gamma }}_{obs,PDMI,WB}$ is consistent for ${\gamma }_{obs}=V^{-1}_{ML}V_{com}$ because ${\widehat{W}}_{PDMI}$ is consistent for $V_{com}$ and ${{\widehat{W}}_{PDMI}+\widehat{B}}_{PDMI}$ is consistent for $V_{ML}$. It follows that ${\widehat{\gamma }}_{mis|PDMI,WB}$ is consistent for ${\gamma }_{mis}$. 

(In the PDMI literature, the fraction of observed information is usually defined a little differently, as $V^{-1}_{PDMI}V_{com}$. Under that definition, the fractions of observed and missing information are consistently estimated by ${\widetilde{\gamma }}_{obs|PDMI,WB}={\widehat{V}}^{-1}_{PDMI}{\widehat{W}}_{PDMI}$ and ${\widetilde{\gamma }}_{mis|PDMI,WB}=I-{\widetilde{\gamma }}_{obs|PDMI,WB}$.)

We can construct a WB confidence interval for scalar $\theta $:
\begin{equation}
\widehat{\theta }_{PDMI}\pm t_{PDMI,WB}\widehat{V}^{1/2}_{PDMI,WB}
\end{equation}

\noindent where $t_{PDMI,WB}$ is a quantile from a $t$ distribution with ${\nu }_{PDMI,WB}$ degrees of freedom ($df$). A simple $df$ estimate is 
\begin{equation}
\widehat{\nu }_{PDMI,WB}=(M-1)\widetilde{\gamma }^{-2}_{mis|PDMI,WB}
\end{equation}

\noindent \cite{Rubin1987}, but this estimate can be highly variable and produce values that are unrealistically large (exceeding the sample size) or unnecessarily small (less than 3). To avoid these problems, we replace ${\widehat{\nu }}_{PDMI,WB}$ with 
\begin{equation} 
\widetilde{\nu }_{PDMI,WB} = max(3,(\widehat{\nu }^{-1}_{PDMI,WB}+\widetilde{\nu }^{-1}_{obs})^{-1})
\end{equation}

\noindent which is bounded below at 3 and above at the $df$ in the observed data, estimated by
\begin{equation}
\widetilde{\nu }_{obs} = {\nu }_{com} \widetilde{\gamma }_{obs|PDMI,WB}\left(\frac{{\nu }_{com}+1}{{\nu }_{com}+3}\right) 
\end{equation}

\noindent where ${\nu }_{com}$ is the $df$ that would be available if the data were complete---e.g., ${\nu }_{com}=N-2$ for a simple linear regression \cite{Barnard1999, vH2015}. If $\theta $ is a vector, we use the same formulas but replace ${\widetilde{\gamma }}_{mis|PDMI,WB}$ with the average of its diagonal elements \cite{Barnard1999}. 

The WB estimators are functions of ${\widehat{B}}_{PDMI}$ and give more weight to ${\widehat{B}}_{PDMI}$ if ${\gamma }_{mis}$ is large. Since ${\widehat{B}}_{MI}$ is imprecise and volatile if $M$ is small, it follows that the WB estimators are imprecise and volatile if $M$ is small and ${\gamma }_{mis}$ is large. The number of imputations $M$ that are needed for stable variance estimates increases quadratically with ${\gamma }_{mis}$ \cite{vH2018}:
\begin{equation}
M = 1 + \frac{1}{2} \left({\gamma }_{mis} / CV \right)^2
\end{equation}

\noindent where $CV$ is the desired coefficient of variation for the SE estimate. For example, if we want $CV=.05$ --- implying that the SE estimate would probably change by less than 5\% if we imputed the data again --- then we should use $M = 1 + 200 {\gamma }_{mis}^2$ imputations --- e.g., just 3 imputations if ${\gamma }_{mis} = .1$ but 51 imputations if ${\gamma }_{mis} = .5$.

\subsection{Under MLMI}
\label{subsection:MLMI_WB}

The WB formulas that are consistent under PDMI are inconsistent under MLMI, and for that reason MLMI has been defined as ``improper.'' But we now present alternative WB estimators that are consistent under MLMI:
\begin{eqnarray}
\label{g_mis_MLMI_WB}
\widehat{\gamma }_{mis|MLMI,WB} &=& {\widehat{W}}^{-1}_{MLMI}{\widehat{B}}_{MLMI} \\
\label{g_obs_MLMI_WB}
\widehat{\gamma }_{obs|MLMI,WB} &=& {I-\widehat{\gamma }}_{mis|MLMI,WB} \\
\label{V_ML_WB}
\widehat{V}_{ML|MLMI,WB} &=& \widehat{W}_{MLMI}\widehat{\gamma }^{-1}_{obs|MLMI,WB} \\
\label{V_MLMI_WB}
\widehat{V}_{MLMI,WB} &=& \widehat{V}_{ML|MLMI,WB}+\frac1{M}\widehat{B}_{MLMI}
\end{eqnarray}

To verify the consistency of these estimators, replace ${\widehat{W}}_{MLMI}$, ${\widehat{B}}_{MLMI}$, and ${\widehat{B}}_{MLMI}/M$ with their estimands: ${\widehat{W}}_{MLMI}$ consistently estimates $V_{com}$, ${\widehat{B}}_{MLMI}$ consistently estimates $V_{com}{\gamma }_{mis}$ (from \eqref{EB_MI}), and ${\widehat{B}}_{MLMI}/M$ consistently estimates $V_{MLMI}-V_{ML}$ (from \eqref{EB_MI_over_M}). 

Although consistent, the WB estimators under MLMI can be imprecise if $M$ is small and ${\gamma }_{mis}$ is large. The imprecision comes again from ${\widehat{B}}_{MLMI}$. In fact, ${\widehat{B}}_{MLMI}$ can be so imprecise that it exceeds ${\widehat{W}}_{MLMI}$. If ${\widehat{B}}_{MLMI}$ is scalar, the fact that it can exceed ${\widehat{W}}_{MLMI}$ means that the estimate ${\widehat{\gamma }}_{mis|MLMI,WB}$ can exceed one, although the estimand ${\gamma }_{mis}$ cannot; therefore the estimates ${\widehat{\gamma }}_{obs|MLMI,WB}$ and ${\widehat{V}}_{ML|MLMI,WB}$ can be negative, although the corresponding estimands must be positive. If ${\widehat{B}}_{MLMI}$ is a matrix, the problem is that the variance estimate ${\widehat{V}}_{ML|MLMI,WB}$ may not be positive definite, although the true variance is. These problems are rare if ${\gamma }_{mis}$ is small, but more common if ${\gamma }_{mis}$ is large and $M$ is small. (See Appendix \ref{appendix:shrinkage_estimator}.)

To increase precision and avoid negative estimates, if ${\widehat{\gamma }}_{mis|MLMI,WB}$ is a scalar we replace it with a shrunken estimator that is guaranteed to take values between 0 and 1:
\begin{equation}
\widetilde{\gamma }_{mis|MLMI,WB}=h(\widehat{\gamma }_{mis|MLMI,WB},M-1)
\end{equation}

\noindent Here the shrinkage function is
\begin{equation} 
\label{shrink_scalar}
h(\widehat{\gamma },\nu)=\frac{\nu }{2}\widehat{\gamma }\frac{{\Gamma}(\frac{\nu -2}{2},\frac{\nu }{2}\widehat{\gamma })}{{\Gamma}(\frac{\nu}{2},\frac{\nu }{2}\widehat{\gamma })}
\end{equation}

\noindent where ${\Gamma}(a,z)$ is the upper incomplete gamma function. This shrinkage function is derived in Appendix \ref{appendix:shrinkage_estimator}.

If $\widehat{\gamma }$ is a matrix, the shrinkage function becomes 
\begin{equation} 
\label{shrink_matrix}
H(\widehat{\gamma }_{mis|MLMI,WB},\nu)=Q\widetilde{\Lambda }Q^{-1}
\end{equation}
\noindent where \textit{Q} is the eigenvector matrix for $\widehat{\gamma }$, and $\widetilde{\mathrm{\Lambda }}$ is a diagonal matrix of eigenvalues, each shrunk by \textit{h}(). This requires that all the eigenvalues are nonzero, which in turn requires that $M$ exceeds the number of rows in $\widehat{\gamma }$.

The shrunken estimator $\widetilde{\gamma }_{mis|MLMI,WB}$ is guaranteed to have eigenvalues between 0 and 1, and the shrunken estimator $\widetilde{V}_{MLMI,WB}$ is guaranteed to be positive definite. In addition, the shrunken variance estimator $\widetilde{V}_{MLMI,WB}$ is less variable than the non-shrunken estimator $\widehat{V}_{MLMI,WB}$. There is more shrinkage if ${\widehat{\gamma }}_{mis|MLMI,WB}$ is large or $M$ is small, and less shrinkage otherwise.

Shrunken estimates of ${\gamma }_{obs}$, $V_{ML}$, and $V_{MLMI}$ can be obtained by substituting $\widetilde{\gamma }_{mis|MLMI,WB}$ for $\widehat{\gamma }_{mis|MLMI,WB}$ in equations \eqref{g_obs_MLMI_WB}, \eqref{V_ML_WB}, and \eqref{V_MLMI_WB}. The shrunken estimates $\widetilde{\gamma }_{obs|MLMI,WB}$, $\widetilde{V}_{ML,WB}$, and $\widetilde{V}_{MLMI,WB}$ are guaranteed to be positive definite; they are also less variable than their non-shrunken counterparts $\widehat{\gamma }_{obs|MLMI,WB}$, $\widehat{V}_{ML,WB}$, and $\widehat{V}_{MLMI,WB}$.

The cost of shrinkage is that the shrunken estimators $\widetilde{\gamma }_{mis|MLMI,WB}$, $\widetilde{V}_{ML,WB}$, and $\widetilde{V}_{MLMI,WB}$ are  biased toward zero (too small on average) if ${\gamma }_{mis}$ is large and $M$ is small relative to ${\gamma }_{mis}$. Table \ref{tab:imputations_for_bias} uses numerical integration (see Appendix \ref{appendix:shrinkage_estimator}) to estimate the number of imputations that are needed to avoid negative bias in ${\widetilde{V}}_{MLMI|WB}$. Ten or fewer imputations suffice if ${\gamma }_{mis}\le .6$, which covers most practical settings. Above ${\gamma }_{mis}>.6$, the number of imputations required by MLMI increases quickly, but may still be practical since MLMI outputs imputations more quickly than PDMI.

\begin{table}
\caption{Number of imputations needed for approximately unbiased shrunken WB estimates under MLMI}
\label{tab:imputations_for_bias}
\begin{tabular}{ c c } \hline 
${\gamma }_{mis}$ & Imputations \\ \hline 
.1 & 2  \\ 
.2 & 2  \\ 
.3 & 2  \\ 
.4 & 3  \\  
.5 & 5  \\ 
.6 & 10  \\  
.7 & 20  \\ 
.8 & 60  \\ 
.9 & 300  \\ \hline
\end{tabular}
\end{table}

If $\theta $ is scalar, we can offer a CI:
\begin{equation}
\widehat{\theta }_{MLMI}\pm t_{MLMI,WB}\widetilde{V}^{1/2}_{MLMI,WB}
\end{equation}
where $t_{MLMI,WB}$ is a quantile from a $t$ distribution whose $df$ are approximated in Appendix \ref{appendix:df}:
\begin{equation}
\label{df_MLMI_WB}
\widehat{\nu}_{MLMI,WB}=\frac{\widetilde{V}^2_{MLMI,WB}}{\frac{{\widetilde{V}}^2_{ML,WB}}{{\widetilde{\nu }}_{ML,WB}}+\frac{(\frac1{M}{\widehat{B}}_{MLMI})^2}{M-1}}
\end{equation}

\noindent where 
\begin{equation}
{\widetilde{\nu }}_{ML,WB} = (M-1)\left(\frac{\widetilde{{\gamma}}_{obs}}{\widetilde{{\gamma}}_{mis}}\right)^2-4
\end{equation}

\noindent Notice that ${\widehat{\nu }}_{MLMI,WB}$ converges to ${\widetilde{\nu }}_{ML,WB}$ as $M$ gets large.

As is the case under PDMI, under MLMI the $df$ estimate can be highly variable and it is helpful to prevent it from getting too high or too low. To accomplish this, we adapt the PDMI formula and replace ${\widehat{\nu }}_{MLMI,WB}$ with 
\begin{equation}
\widetilde{\nu }_{MLMI,WB}=max(3,\widehat{\nu }^{-1}_{MLMI,WB}+\widetilde{\nu }^{-1}_{obs})^{-1})
\end{equation}

\noindent where ${\widetilde{\nu }}_{obs}={\nu }_{com}{\widetilde{\gamma }}_{obs|MLMI,WB}(\frac{{\nu }_{com}+3}{{\nu }_{com}+1})$ estimates the $df$ in the observed data. 

If $\theta $ is a vector, we use the same $df$ formulas but replace ${\widetilde{V}}_{ML,WB}$, ${\widetilde{V}}_{MLMI,WB}$, and ${\widehat{B}}_{MLMI}$ with their diagonal elements and replace $\widetilde{\gamma }_{obs \mid MLMI,WB}$ and ${\widetilde{\gamma }}_{mis \mid MLMI,WB}$ with the average of their diagonal elements.

\section{ Score-based (SB) variance estimation} \label{section:SB_variance_estimation}

As an alternative to WB variance estimation, Wang and Robins \cite{Wang1998} proposed a \textit{score-based} (SB) variance estimator, which used the \textit{score function}, defined using the contribution of each case to the gradient of the log likelihood. Their formula was somewhat complicated, and we derive a simpler alternative, which Appendix \ref{appendix:SB_estimator_WR} shows is equivalent in large samples. The same SB formulas apply under PDMI or MLMI.

The SB formulas are less often usable than the WB formulas, because the score function is often unavailable to the user. The user typically does not see the score function when they maximize the likelihood, and some common estimation techniques, such as least squares, do not maximize the likelihood explicitly, but obtain equivalent estimates by other means. In addition, the SB formula assumes independently and identically distributed (iid) observations, which the WB formulas do not assume.

Here is a derivation of our SB formula. Let $S_{com}={\nabla }ln L ({\theta} \mid Y_{com})$ be the complete-data score that would be available with complete data, and let $S_{obs}={\nabla }ln L ({\theta} \mid Y_{obs})$ be the observed-data score that is available given the observed data. Both scores have expectations of zero. The variance of the complete-data score is the complete-data information $V^{-1}_{com}=V(S_{com})$. The variance of the observed-data score is the observed-data information $V(S_{obs})=V^{-1}_{ML}$.

In iid data, each observation makes an equally weighted contribution to the score. In complete data, the score can be expressed as the sum $S_{com}=\sum^N_{i=1}{s_{com,i}}$, where each summand $s_{com,i}={\nabla }ln L ({\theta} \mid y_{com,i})$ is a function of the parameters $\theta $ and the values $y_{com,i}$ of the complete data in observation \textit{i}. We can think of $s_{com,i}$ as a variable with a different value in each observation. Then $s_{com,i}$ has an expectation of zero and a variance of $V(s_{com,i})=V^{-1}_{com}N^{-1}$.

We can estimate $s_{com,i}$ using MI data. For observation $i$ in SI dataset $m$, the estimate is 
\begin{equation}
\widehat{s}_{com,i,m}={\nabla } ln L(\widehat{\theta }_{MI}\mid y_{SI,i,m})
\end{equation}

\noindent and the variance (over $i$) of ${\widehat{s}}_{com,i,m}$ consistently estimates $V^{-1}_{com}N^{-1}$. In addition, ${\widehat{s}}_{com,i,m}$ can be split into random effects components. One component lies between observations, and the other component lies within observations --- i.e., between different imputations of the same observation:
\begin{equation}
\widehat{s}_{com,i,m}=s_{\infty I,i}+d_{SI,m,i}
\end{equation}

The between-observation component $s_{\infty I,i}$ is the average of ${\widehat{s}}_{com,m,i}$ across the infinite population of imputations; in large samples, $s_{\infty I,i}$ is equivalent to $s_{obs,i}=\mathrm{\nabla }{\mathrm{ln} L\ }(\theta |y_{obs,i})$, which is the contribution of case \textit{i} to $S_{obs}$. The within-observation component $d_{SI,m,i}$ is the imputation-specific departure of ${\widehat{s}}_{com,m,i}$ from the average $s_{\infty I,i}$. The components have expectations of zero and asymptotic variances (over $i$) of
\begin{eqnarray}
V(\widehat{s}_{com,m,i}) \xrightarrow{N \to \infty} \frac1{N}V^{-1}_{com} \\
V(s_{\infty I,i})  \xrightarrow{N \to \infty} \frac1{N}V^{-1}_{ML} \\
V(d_{SI,m,i})  \xrightarrow{N \to \infty} \frac1{N}V^{-1}_{mis}  
\end{eqnarray}

We can estimate the variance components using MANOVA, and multiply the variance estimates by $N$ to obtain estimators of $V^{-1}_{com}$, $V^{-1}_{mis}$, and $V^{-1}_{ML}$: 
\begin{eqnarray}
\widehat{V}^{-1}_{com|SB} &=& \frac{SST}{M}=\frac1{M}\sum^M_{m=1} \sum^N_{i=1} \widehat{s}^{\otimes 2}_{com,m,i} \\
\widehat{V}^{-1}_{mis|SB} &=& \frac{SSW}{M-1}=\frac1{M-1 }\sum^M_{m=1} \sum^N_{i=1} (\widehat{s}_{com,m,i}-\overline{s}_{com,i})^{\otimes 2} \\
\widehat{V}^{-1}_{ML|SB} &=& \widehat{V}^{-1}_{com|SB}-\widehat{V}^{-1}_{mis|SB}
\end{eqnarray}

\noindent where ${\overline{s}}_{com,i}=M^{-1}\sum^M_{m=1}{{\widehat{s}}_{com,m,i}}$, and \textit{SST} and \textit{SSW} are the total and within sums of squares. We can use these results to derive estimators that are consistent for ${\gamma }_{mis}$ and $\ {\gamma }_{obs}$:
\begin{eqnarray}
\widehat{\gamma }_{mis|SB} &=& \widehat{V}^{-1}_{mis|SB}\widehat{V}_{com|SB} \\
\widehat{\gamma }_{obs|SB} &=& I-\widehat{\gamma }_{mis}
\end{eqnarray}

It occasionally happens that ${\widehat{V}}_{ML|SB}$ and ${\widehat{\gamma }}_{obs|SB}$ will fail to be positive definite, especially if $M$ is small and ${\gamma }_{mis}$ is large. This happens when some of the eigenvalues of ${\widehat{\gamma }}_{mis|SB}$ exceed 1. To guarantee positive definiteness, we shrink the estimators as follows:
\begin{eqnarray}
\widetilde{\gamma }_{mis|SB} &=& H(\widehat{\gamma }_{mis|SB},(M-1)N) \\
\widetilde{\gamma }_{obs|SB} &=& I-\widetilde{\gamma }_{mis|SB} \\
\widetilde{V}_{ML|SB} &=& \widehat{V}_{com|SB} \widetilde{\gamma }^{-1}_{obs|SB}
\end{eqnarray}

\noindent where the shrinkage function ${H}()$ was defined in \eqref{shrink_matrix}.

Then an SB estimator for the variance of an MI point estimate is
\begin{equation}
\widetilde{V}_{MI|SB}=\widetilde{V}_{ML|SB}+\frac1{M}\widehat{B}_{MI}
\label{V_MI|SB}
\end{equation}

\noindent $\widetilde{V}_{MI|SB}$ consistently estimates $V_{MI}$ because $\widetilde{V}_{ML|SB}$ consistently estimates $V_{ML}$ and $\widehat{B}_{MI}/M$ consistently estimates $V_{MI}-V_{ML}$. 

An SB CI for scalar $\theta$ is
\begin{equation} 
\widehat{\theta }_{BMI}\pm t_{SB}\widetilde{V}^{1/2}_{MI|SB}
\end{equation}

\noindent where $t_{SB}$ is a quantile from a $t$ distribution with $df={\nu }_{SB}$, which is the $df$ of $\widetilde{V}_{MI|SB}$ 

It remains only to estimate ${\nu }_{SB}$. Since ${\widehat{B}}_{MI}$ has $df$=$M$--1 and ${\widetilde{V}}_{ML|SB}$ may be assumed to have $df$ no less than ${\widetilde{\nu }}_{obs|SB}={\nu }_{com}{\widetilde{\gamma }}_{obs|SB}(\frac{{\nu }_{com}+3}{{\nu }_{com}+1})$, a Satterthwaite approximation for ${\nu }_{SB}$ is 
\begin{equation}
\widehat{\nu }_{SB} = \frac {\widetilde{V}_{MI|SB}^2 }
{ \frac {\widetilde{V}_{ML|SB}^2} {\widetilde{\nu}_{obs|SB}} + \frac {\left(\frac1{M} \widehat{B}_{MI}\right)^2}{M-1} }
\end{equation}

\noindent which is very close to ${\widetilde{\nu }}_{obs|SB}$ unless $M$ is very small. If $N$ and $M$ are large then $\widehat{\nu }_{SB}$ approaches
\begin{equation}
\widehat{\nu }_{SB} \xrightarrow{N,M \to \infty}
\begin{cases}

(M-1)\left(\frac{M}{\gamma_{obs}\gamma_{mis}}\right)^2 \text{under MLMI} \\    
(M-1)\left(\frac{M}{\gamma_{mis}}\right)^2 \text{under PDMI}
\end{cases}
\end{equation}

\noindent So that asymptotic degrees of freedom are larger under MLMI than under PDMI.

\section{Conditions for consistency of WB and SB variance estimates}
\label{section:assumptions}
The derivations of the WB and SB variance formulas make certain assumptions. If those assumptions are not met, then the resulting variance estimates are not necessarily consistent.

\subsection{Compatible and correctly specified models} 
The WB and SB variance formulas assume that that the same model, with the same parameters $\theta$, is used for imputation and analysis. The formulas also assume that this model is correctly specified \cite{Wang1998}. In practice, though, the analysis model is often different from the imputation model, and one or both models may be misspecified. 

When the analysis and estimation models are different, WB and SB formulas still yield consistent variance estimates if both models are ``compatible'' with some \textit{common model}, and that common model is correctly specified \cite{Bartlett2015}. For example, later, in the simulations, we will consider the situation where the imputation model is a linear regression of $Y$ on $X$ and the analysis model is a linear regression of $X$ on $Y$. If both regression models have normal residuals, then both are compatible with a common model in which $(X,Y)$ are bivariate normal.

If the imputation and analysis models are different, but compatible and correct, then the derivations of the WB and SB variance formulas are valid provided we redefine the parameter vector $\theta$ to include all the parameters of the common model, and not just the parameters of the analysis model. 

How much do the extra parameters in $\theta$ matter for the variance formulas? It depends which formula you use, as we discuss next.

\subsection{Which variance formulas must include all parameters of the common model?}
\label{section:matrix_vs_scalar}

Under PDMI, a nice property of the WB formula \eqref{V_PDMI} is that it uses only addition; it is a weighted sum of ${\widehat{W}}_{PDMI}$ and ${\widehat{B}}_{PDMI}$. As a result, the diagonal components --- i.e., the squared standard error estimates --- in ${\widehat{V}}_{PDMI,WB}$ depend only on the corresponding diagonal components of ${\widehat{W}}_{PDMI}$ and ${\widehat{B}}_{PDMI}$. 

This means that the PDMI WB formula can be applied to any submatrix of ${\widehat{W}}_{PDMI}$ and ${\widehat{B}}_{PDMI}$ and the resulting standard error estimates will not change. In other words, you can apply the PDMI WB formula to any subset of the parameters in $\theta$. In fact, you can apply the PDMI WB formula, in scalar form, to each component of $\theta$, and the standard error estimates will still be the same.

Because of this property, the standard error estimates that come from the PDMI WB formula do not change when you include parameters that are not in the analysis model but are in the common model. You can safely neglect those extra parameters; you don't even have to know what they are. When using the PDMI WB formula, you can limit your attention to the parameters in the analysis model. The resulting standard errors will be consistent if the analysis and imputation models are correct and compatible.

Under MLMI, unfortunately, the WB formula \eqref{V_MLMI} does not have the same property. It must be applied in matrix form, and if the imputation and analysis models are not the same, it must be applied to the whole parameter vector $\theta$ of the common model --- and not just selected components, such as the parameters of the analysis model. Because the MLMI WB formula involves matrix multiplication, the diagonal elements of ${\widehat{V}}_{MLMI,WB}$ can be affected by the off-diagonal elements of ${\widehat{B}}_{MLMI,WB}$ and ${\widehat{W}}_{MLMI,WB}$.  

The SB variance formula \eqref{V_MI|SB} has the same issue. It must be applied in matrix form, and if the imputation and analysis models are not the same, it must be applied to the whole parameter vector $\theta$ of the common model. That's because the SB variance formula uses matrix multiplication, so the off-diagonal elements of $\widehat{V}_{mis|SB}$ and $\widehat{V}_{com|SB}$ can affect the diagonal elements of $\widetilde{V}_{MI|SB}$. We will return to this issue in the simulations.

\section{ Bootstrap variance estimation}
\label{section:BMI_variance}

Unlike the WB and SB formulas, bootstrapped MI (BMI) offers consistent variance estimates and confidence intervals with nominal coverage even when the imputation and analysis models are incompatible, or even incorrect. BMI variance formulas are straightforward and do not require matrix calculations or inclusion of parameters beyond those in the analysis model. The same BMI variance formulas are consistent under BMLMI and under BPDMI. 

Remember that the individual estimates ${\widehat{\theta }}_{bd}$ fit this random effects model (\ref{bootstrap_re_model}):
\begin{equation}
\widehat{\theta }_{bd}=\widehat{\theta }_{ML} +e_b+e_{bd} 
\end{equation}
\noindent The variance components are $V_{ML}=V(e_b)$ and $V_{BD}=V(e_{bd})=V_{SI}-V_{ML}$. To estimate the variance components, we fit the model using ANOVA (or MANOVA) and use mean squared formulas:
\begin{eqnarray} 
\widehat{V}_{BD|BMI} &=& MSW \\
\widehat{V}_{ML|BMI} &=& \frac{MSB - MSW} {M}
\end{eqnarray}

\noindent where $MSB$ is the mean square between the bootstrapped datasets, with $df=B-1$, and $MSW$ is the mean square within the bootstrapped datasets and between the imputed datasets, with $df=B(D-1)$. Then $V_{BMI}=V(\widehat{\theta}_{BMI})$ is estimated by
\begin{equation}
\widehat{V}_{BMI} = \widehat{V}_{ML|BMI}\left(1+\frac{1}{B}\right)+\frac{\widehat{V}_{BD|BMI}}{BD}
\label{Vhat_BMI}
\end{equation}
This estimate is consistent because it replaces each component of the true variance in \eqref{V_BMI} with a consistent estimate.

$\widehat{V}_{BMI}$ can be re-expressed as a weighted sum of independent mean squares
\begin{equation}
\widehat{V}_{BMI} = \frac{1}{M} \left(MSB\left(1+\frac{1}{B}\right) - MSW\right)
\end{equation}

\noindent which according to the Satterthwaite approximation has the following $df$:
\begin{equation}
\label{df_BMI}
{\widehat{\nu }}_{BMI}=\frac{
\left(MSB(B+1) - MSW\left(B\right)\right)^2
}{
\frac{MSB^2(B+1)^2}{B-1}
+\frac{MSW^2 B}{D-1}
}
\end{equation}

\noindent If $D=2$, as we recommended earlier, then as $B$ gets larger, ${\widehat{\nu }}_{BMI}$ approaches
\begin{equation}
\lim_{B\to\infty,D=2}{\widehat{\nu }}_{BMI}=B\left(1-2\frac{MSB\times MSW}{MSB^2+MSW^2}\right)
\end{equation}
which is just a little smaller than $B$ if the fraction of missing information is not too large. 

If $\theta$ is a scalar parameter, then a confidence interval is
\begin{equation}
\label{bootstrap_t_interval}
\widehat{\theta }_{BMI}\pm t_{BMI}\widehat{V}^{1/2}_{BMI}
\end{equation}

\noindent where $t_{BMI}$ is a quantile from a $t$ distribution with $df={\widehat{\nu }}_{BMI}$. Our $df$ and CI formulas assume a scalar $\theta$. If $\theta$ is a vector, then the same formulas apply separately to each scalar component. 

Notice that BMI variance estimation does not require an estimate of the complete-data variance $V_{com}$. But an estimate of $V_{com}$ is necessary to estimate the fractions of observed and missing information. To get those estimates, start with a consistent estimate ${\widehat{V}}_{com,bd}$ obtained by analyzing each of the bootstrapped-then-imputed datasets as though it were complete. The average of the ${\widehat{V}}_{com,bd}$ is a consistent estimate of $V_{com}$:
\begin{equation}
\widehat{V}_{com|BMI}=\frac{1}{BD}\sum^B_{b=1}{\sum^M_{m=1}\widehat{V}_{com,bd}} 
\end{equation}

It follows that 
\begin{eqnarray}
\widehat{\gamma}_{obs,BMI} &=& \widehat{V}^{-1}_{ML|BMI}\widehat{V}_{com|BMI} \\
\widehat{\gamma}_{mis,BMI} &=& I-\widehat{\gamma}_{obs,BMI}
\end{eqnarray}

\noindent are consistent estimators for the fractions of observed and missing information. 

\subsection{How many imputations are needed for variance estimation?} \label{df_variance_section}

Table \ref{tab:point_estimates_imputations} gave the number of imputations that were needed for relatively efficient point estimates. But more imputations may be needed to estimate variances and CIs. At a minimum, a variance estimate should be approximately unbiased if $N$ and $M$ are large. Most of our variance estimates will have little or no bias even if $M$ is small. The one exception is the WB variance estimate under MLMI, and Table \ref{tab:imputations_for_bias} gave the number of imputations that were needed to reduce its bias to a negligible level.

But we often want more from a variance estimate than lack of bias. We also want variance estimates to be \textit{replicable} in the sense that approximately the same variance estimate would be obtained if the data were imputed again, or bootstrapped and imputed again. And we want the confidence interval derived from the variance estimate to be reasonably short,

The $df$ of the variance estimate is a useful guide to these properties. The coefficient of variation for an SE estimate is approximately $\sqrt{1/(2 df)}$ \cite{vH2018}. So at $df=25$ an SE estimate would likely change by about 14\%, and at $df=100$ an SE estimate would likely change by about 7\%, if the data were multiply imputed again --- or bootstrapped and imputed again under BMI. 

Table \ref{tab:imputations_for_df} gives the number of imputations $M$, or bootstrap samples and imputations $BD$, that are needed for different variance estimates to have at least 25, or at least 100, degrees of freedom.

\begin{table}
\caption{Number of imputations needed for variance estimates with specified degrees of freedom}
\label{tab:imputations_for_df}
\begin{subtable}{\textwidth}
            \centering
            \caption{Imputations needed for $df \geq 25$.}
\begin{tabular}{c c c c c c c } \hline 
& \multicolumn{2}{c}{Score-based variance}&  \multicolumn{2}{c}{Within-between variance} & \multicolumn{2}{c}{Boostrapped variance} \\
 \cline{2-7}
$\gamma_{mis}$ & PDMI & MLMI & PDMI & MLMI & PDMI & MLMI \\ \hline
.1&2&2&2&             2 &52&52 \\
.2&2&2&2&             3 &52&52 \\
.3&2&2&4&             7 &52&52 \\
.4&2&2&4&           14 &52&52 \\
.5&3&2&8&           30 &52&52 \\
.6&3&2&10&           67 &52&52 \\
.7&3&2&14&         159 &52&52 \\
.8&3&2&17&         465 &52&52 \\
.9&4&2&22&     2,350 &52&52 \\ \hline
\end{tabular}
\end{subtable}

\begin{subtable}{\textwidth}
            \centering
            \caption{Imputations needed for $df \geq 100$.}
\begin{tabular}{c c c c c c c } \hline 
& \multicolumn{2}{c}{Score-based variance}&  \multicolumn{2}{c}{Within-between variance} & \multicolumn{2}{c}{Boostrapped variance} \\
 \cline{2-7}
$\gamma_{mis}$ & PDMI & MLMI & PDMI & MLMI & PDMI & MLMI \\ \hline
.1&2&2&2&             3 &202&202 \\
.2&2&2&5&             8 &202&202 \\
.3&3&3&10&          21 &202&202 \\
.4&3&3&17&          48 &202&202 \\
.5&4&3&26&        105 &202&202 \\
.6&4&3&37&        235 &202&202 \\
.7&5&3&50&        568 &202&202 \\
.8&5&2&65&     1,665 &202&202 \\
.9&5&2&82&     8,425 &202&202 \\
 \hline
\end{tabular}
\end{subtable}

\textit{Note}. For MI, the number of imputations is $M$. For BMI, the number of imputations is $BD$, where $B$ is the number of bootstrap samples and $D$ is the number of imputations per bootstrap sample.
\end{table}

The SB variance estimates have remarkably modest needs, requiring 5 imputations or less even when the fraction of missing information is very large. Unfortunately, SB variance estimates are often unavailable in practice, since they require a score function which the analyst may not have. 

The WB estimates need few imputations when the fraction of missing information is small, but require more and more imputations as the fraction of missing information grows, especially under MLMI. 

The BMI variance estimators require $BD=2(df+1)$ imputations, regardless of the fraction of missing information. Under PDMI, BMI needs more imputations than the WB estimator even when the fraction of missing information is as large as .9. Under MLMI, BMI needs more imputation than the WB estimator if the fraction of missing information is less than .6, but BMI needs fewer imputations than the WB estimator if the fraction of missing information is .6 or greater. Under MLMI, therefore, if the fraction of missing information is large there is no reason to use the WB estimator when the fraction of missing information is large; instead, switch to BMI.

Remember that BMI variance estimator is consistent under circumstances when the WB and SB estimators may be inconsistent. Therefore BMI should be preferred when there is enough time to produce the number of imputations that it requires. And more imputations can be produced more quickly using MLMI than using PDMI.

\section{Software}
The second author implemented all the methods described here and published them in new R packages called \textit{mlmi} and \textit{bootImpute} \cite{Bartlett2019a,Bartlett2019b}. 

The \textit{mlmi} package implements MLMI and PDMI versions of four different imputation models: (1) normal linear regression of one incomplete variable on one or more complete variables, (2) the multivariate normal model for data with several incomplete continuous variables, (3) the log-linear model for data with several incomplete categorical variables, and (4) the general location for a ``mix'' of categorical and continuous variables. The general location model can be described as a multivariate normal model whose mean is conditioned on a log-linear model of the categorical variables \cite{Schafer1997}. The \textit{mlmi} package also implements the SB formulas and WB formulas that are appropriate for data imputed using MLMI and PDMI. When using the SB formulas, the user must specify the score funcation.

The \textit{bootImpute} package implements bootstrapped MI and the formulas that are used to calculate standard errors and confidence intervals from bootstrapped MI data. The \textit{bootImpute} package can be used with any imputation function, using either MLMI or PDMI. The \textit{bootImpute} package includes functions that integrate it with the popular $mice$ package \cite{vanBuuren2018b}, which imputes missing values using a set of regression model, and the $smcfcs$ package \cite{Bartlett2019c}, which modifies the $mice$ approach to ensure that the imputation and analysis models are compatible. 

The second author used these R packages to carry out simulations and an applied data analysis in R. The simulation and analysis code resides in a github repository at https://github.com/jwb133/mlmiPaper. Some of the simulations were replicated independently by the first author in SAS.

\section{Simulations}

In this section, we use simulation to compare the properties of MLMI and PDMI, with and without the bootstrap.

\subsection{Design}
\label{section:Design}

We simulated $N$ rows of standard bivariate normal data $(X,Y)$ 
\begin{equation}
\begin{bmatrix}
X \\
Y
\end{bmatrix}
\sim
N_2\left(\begin{bmatrix}
\mu_X \\
\mu_Y
\end{bmatrix},
\begin{bmatrix}
\sigma_X^2 & \rho\sigma_X\sigma_Y \\
\rho\sigma_X\sigma_Y & \sigma_Y^2
\end{bmatrix}
\right)
\end{equation}

\noindent with correlation $\rho =.5$, means $\mu_X=\mu_Y=0$, and variances $\sigma_X^2=\sigma_Y^2=1$. The data fit a linear regression of $Y$ on $X$, or of $X$ on $Y$:
\begin{eqnarray}
Y=\alpha_{Y} +\beta_{Y.X} X + e_{Y.X}, \text{where } e_{Y.X} \sim N(0,\sigma^2_{Y.X}) \\
X=\alpha_{X} +\beta_{X.Y} Y + e_{X.Y}, \text{where } e_{X.Y} \sim N(0,\sigma^2_{X.Y}) \end{eqnarray}

\noindent The parameters of both regressions have the same values: $\alpha_{Y}=\alpha_{X}=0$, $\beta_{Y.X}=\beta_{X.Y} =\rho $, and ${\sigma^2_{Y.X}}={\sigma^2_{X.Y}}=1-{\rho }^2$.

We then deleted some fraction---either $p=.25$ or $p=.5$---of $Y$ values in one of two patterns:
\begin{itemize}
\item  \textit{Missing completely at random (MCAR)}. Each $Y$ value has an equal probability \textit{p} of being deleted.

\item  \textit{Missing at random (MAR)}. $Y$ is more likely to be deleted if $X$ is large. In particular, $Y$ is deleted with probability $2p \Phi(X)$, where $\Phi$ is the standard normal CDF. 
\end{itemize}

\noindent For a given value of $p$, the fraction of observed information ${\gamma }_{obs}$ was lower under MAR than under MCAR.

We imputed missing $Y$ values using the following imputation model: 
\begin{equation}
Y_i=\widehat{\alpha }_{Y.X}+\widehat{\beta }_{Y.X}X_i+e_i, \text{where } e_i\sim N(0,{\widehat{\sigma}_{Y.X}}^2). 
\end{equation}

\noindent The parameter estimates $\widehat{\alpha }_{Y.X},\widehat{\beta }_{Y.X},{\widehat{\sigma}_{Y.X}}^2$ were ML estimates under MLMI and PD estimates under PDMI. In this simple setting, with $X$ complete and $Y$ MAR or MCAR, we could get ML and PD estimates non-iteratively. We got ML estimates $\widehat{\alpha}_{Y.X,ML}$, $\widehat{\beta}_{Y.X,ML}$ by OLS regression of $Y$ on $X$ in the $n_Y$ cases with $Y$ observed; then we calculated the ML estimate $\widehat{\sigma}_{Y.X,ML}^2$ by dividing the residual sum of squares by $n_Y$ \cite{Anderson1957}. We got PD estimates by drawing from the following distributions \cite{Kim2004}:
\begin{equation}
\widehat{\sigma}_{Y.X,PD}^2 \sim \frac{n_Y}{U}\widehat{\sigma}_{Y.X,ML}^2 
\end{equation}
\begin{equation}
\begin{bmatrix}
\widehat{\alpha}_{Y.X,PD} \\
\widehat{\beta}_{Y.X,PD}
\end{bmatrix}
\sim
N_2\left(\begin{bmatrix}
\widehat{\alpha}_{Y.X,ML} \\
\widehat{\beta}_{Y.X,ML}
\end{bmatrix},\widehat{V}_{ML}\frac{\widehat{\sigma}_{Y.X,PD}^2}{\widehat{\sigma}_{Y.X,ML}^2}\right)
\end{equation}

\noindent where $N_2()$ is the bivariate normal distribution, $\widehat{V}_{ML}$ is the estimated variance of the ML estimates $\widehat{\alpha}_{Y.X,ML}$, $\widehat{\beta}_{Y.X,ML}$, and $U$ is a chi-squared random variable with degrees of freedom $n_Y-2+\nu_{prior}$ . Here $\nu_{prior}$ is the prior degrees of freedom, which we set conventionally to 0, although 2 is a better choice \cite{Kim2004, vH2013}.

In the imputed data, we regressed the incomplete variable $Y$ on the complete variable $X$, and then reversed the regression, regressing $X$ on $Y$. Using formulas derived in previous sections, we calculated regression point estimates and their estimated covariance matrix, along with standard error estimates and confidence intervals. 

When using matrix formulas to calculate the covariance matrix of the regression estimates the question arose how large a matrix we must use. As discussed in section \ref{section:matrix_vs_scalar}, the answer depends on whether the imputation model and the analysis model were the same:
\begin{itemize}
    \item When the analysis regressed $Y$ on $X$, the analysis model was the same as the imputation model, and we could limit calculations to the $2\times2$ covariance matrix of the parameter estimates  $(\widehat{\alpha}_{Y.X},\widehat{\beta}_{Y.X})$. (We could have used a  $3\times3$ matrix that included $\widehat{\sigma^2}_{X.Y}$, but this was not necessary because $\widehat{\sigma^2}_{X.Y}$ is uncorrelated with $(\widehat{\alpha}_{Y.X},\widehat{\beta}_{Y.X})$.)
    \item When the analysis regressed $X$ on $Y$, the analysis model differed from the imputation model, but both were compatible with a common bivariate normal model for $(X,Y)$. So the matrix calculations must use a $5\times5$ matrix that includes the covariances among the 5 estimated parameters of the bivariate normal distribution. There are several ways to parameterize the bivariate normal distribution. We chose the parameterization $(\alpha_{X.Y},\beta_{X.Y},\sigma^2_{X.Y},\mu_Y,\sigma^2_Y)$ because it includes the parameters for the regression of $X$ on $Y$.\footnote{This parameterization results from factoring the bivariate normal distribution as $N_2(X,Y)=f(Y)f(X|Y)$.}
\end{itemize}

As noted in section \ref{section:matrix_vs_scalar}, the size of the covariance matrix matters only for the SB formula and the MLMI WB formula. When using the PDMI WB formula or the bootstrap formula, the size of the matrix does not affect estimated standard errors or confidence intervals.

We ran the simulation at two different sample sizes: $N=100$ and 500. At each sample size, we used $M=10$, 50, or 200 imputations. When using the bootstrap, we set $BD$=50 or 200, where $B$=25 or 100 is the number of bootstrap samples, and $D=2$ is the number of imputations per bootstrap sample.\footnote{We considered a condition with $M=5$ imputations, but decided against it since some $B_{MI}$ matrices are $5\times5$ and would not be positive definite with $M=5$. We also decided againsta condition $B=5$ bootstrap samples, as the resulting variance estimates would have only about 4 degrees of freedom.} We replicated each simulated condition 10,000 times, so that the coverage of 95\% confidence intervals was estimated within a standard error of 0.2\%. 

\subsection{Results}

In presenting simulation results, we focus on the regression slope $\beta_{Y.X}$ or $\beta_{X.Y}$, though we got similar results, not shown, for the intercept. We summarized the accuracy of point estimates using the percent root mean squared error (RMSE) -- i.e., the RMSE of a scalar parameter estimate $\widehat{\beta}_{Y.X}$ or $\widehat{\beta}_{X.Y}$ expressed as a percentage of the true parameter value $\beta_{Y.X}$ or $\beta_{X.Y}$. In the regression of the incomplete $Y$ on the complete $X$, the estimate $\widehat{\beta}_{Y.X}$ is unbiased, so the RMSE reflects  variability only. In the regression of $X$ on $Y$, though, the estimate $\widehat{\beta}_{Y.X}$ is biased in small samples \cite{vH2015}, so the RMSE reflects bias as well as variability.

\subsubsection{Regression of $Y$ on $X$}

We first regressed $Y$ on $X$. Here the analysis model is the same as the imputation model, so all matrix calculations are limited to the two model parameters  $({\alpha}_{Y.X},{\beta}_{Y.X})$. See section \ref{section:Design} for explanation. 

\begin{table}
\caption{Estimating the slope of Y on X.}
\label{tab:Y_on_X}
\begin{subtable}{\textwidth}
\caption{Percent root mean square error of point estimates.}
            \label{tab:Y_on_X_RMSE}
\begin{tabular}{ c c c c c c c}
\hline
\multicolumn{2}{c}{Missing}&&\multicolumn{2}{c}{Repeated MI}&\multicolumn{2}{c}{Bootstrapped MI} \\
 \cline{1-2} \cline{4-7} 
\%&Pattern&Imputations&PDMI&MLMI&PDMI&MLMI \\
\hline
25 & MCAR & 10 & 9.1 & 9.1 & & \tabularnewline
 &  & 50 & 9.0 & 9.0 & 9.2 & 9.2\tabularnewline
 &  & 200 & 9.0 & 9.0 & 9.0 & 9.0\tabularnewline
 & MAR & 10 & 9.3 & 9.3 & & \tabularnewline
 &  & 50 & 9.2 & 9.2 & 9.4 & 9.4\tabularnewline
 &  & 200 & 9.2 & 9.2 & 9.3 & 9.2\tabularnewline
50 & MCAR & 10 & 11.3 & 11.2 & & \tabularnewline
 &  & 50 & 11.1 & 11.1 & 11.3 & 11.3\tabularnewline
 &  & 200 & 11.1 & 11.1 & 11.2 & 11.2\tabularnewline
 & MAR & 10 & 13.8 & 13.5 & & \tabularnewline
 &  & 50 & 13.7 & 13.6 & 13.9 & 13.8\tabularnewline
 &  & 200 & 13.5 & 13.5 & 13.5 & 13.5\tabularnewline
\hline
\end{tabular}
\end{subtable}

\bigskip
\begin{subtable}{\textwidth}
            \caption{Mean length of 95\% confidence intervals (CIs). \\ (Parentheses enclose \% departure from 95\% coverage.)}
                        \label{tab:Y_on_X_CI}
\begin{tabular}{ c c c   c c c   c c c}
\hline
\multicolumn{2}{c}{}& &\multicolumn{4}{c}{Repeated MI}  &\multicolumn{2}{c}{ } \\
\cline{4-7}
\multicolumn{2}{c}{Missing}& &\multicolumn{2}{c}{Score-based CIs} &\multicolumn{2}{c}{Within-between CIs} &\multicolumn{2}{c}{Bootstrapped CIs} \\
\cline{1-2} \cline{4-9}
\% & Pattern & Imputations &PDMI &MLMI &PDMI &MLMI &PDMI &MLMI \\
\hline
25 & MCAR & 10 & 0.18 (0.3) & 0.18 (0.3) & 0.18 (0.1) & 0.19 (1.3) &  & \tabularnewline
 &  & 50 & 0.18 (0.5) & 0.18 (0.3) & 0.18 (0.1) & 0.18 (0.3) & 0.19 (-0.3) & 0.19 (-0.4)\tabularnewline
 &  & 200 & 0.18 (0.1) & 0.18 (0.1) & 0.18 (0.1) & 0.18 (0.1) & 0.18 (-0.1) & 0.18 (-0.2)\tabularnewline
 & MAR & 10 & 0.19 (0.4) & 0.18 (0.3) & 0.18 (0.1) & 0.19 (1.3) &  & \tabularnewline
 &  & 50 & 0.18 (0.0) & 0.18 (0.2) & 0.18 (-0.1) & 0.18 (0.2) & 0.19 (-0.2) & 0.19 (-0.3)\tabularnewline
 &  & 200 & 0.18 (0.1) & 0.18 (0.1) & 0.18 (-0.1) & 0.18 (0.0) & 0.18 (-0.4) & 0.18 (-0.2)\tabularnewline
50 & MCAR & 10 & 0.23 (0.6) & 0.22 (0.4) & 0.23 (0.1) & 0.30 (2.1) &  & \tabularnewline
 &  & 50 & 0.22 (0.6) & 0.22 (0.6) & 0.22 (0.1) & 0.24 (1.3) & 0.24 (0.3) & 0.23 (0.0)\tabularnewline
 &  & 200 & 0.22 (0.4) & 0.22 (0.4) & 0.22 (0.0) & 0.22 (0.2) & 0.22 (-0.4) & 0.22 (-0.4)\tabularnewline
 & MAR & 10 & 0.28 (0.7) & 0.27 (0.4) & 0.29 (0.1) & 0.28 (-1.7) & & \tabularnewline
 &  & 50 & 0.27 (0.1) & 0.27 (-0.2) & 0.27 (-0.2) & 0.27 (-1.3) & 0.29 (0.2) & 0.28 (-0.4)\tabularnewline
 &  & 200 & 0.27 (0.1) & 0.27 (0.0) & 0.26 (-0.2) & 0.28 (-0.3) & 0.27 (-0.4) & 0.26 (-0.3)\tabularnewline
\hline
\end{tabular}
\end{subtable}
\medskip
\end{table}

Table \ref{tab:Y_on_X_RMSE} gives the percent RMSE for point estimates of the slope $\beta_{Y.X}$. The RMSE is slightly smaller under repeated MI than under bootstrapped MI, and slightly smaller under MLMI than under PDMI. But most differences in RMSE are very small, even when there is little information or few imputations. For example, even with 10 imputations and 50 percent of values MAR, the RMSE is only 2 percent smaller under MLMI than under PDMI. 

Table \ref{tab:Y_on_X_CI} gives the mean length of nominal 95\% CIs, along with their departure from 95\% coverage. Bootstrapped and SB CIs come within 0.5\% of nominal coverage. They are shorter under MLMI than under PDMI, but the difference is negligible and vanishes as the fraction of missing information gets small or the number of the imputations gets large. 

WB CIs have more accurate coverage under PDMI than under MLMI. They come within 0.2\% of nominal coverage under PDMI, but can drift as far as 2\% above or below nominal coverage under MLMI. Coverage improves with more information or more imputations. Under most conditions, WB CIs are slightly longer, with higher coverage, under MLMI than under PDMI, but with more missing information WB CIs can be shorter under MLMI because of the shrinkage function in equation \ref{shrink_scalar}.

\subsubsection{Regression of $X$ on $Y$}
We next regressed $X$ on $Y$. Since the imputation model is a regression of $Y$ on $X$, the imputation and analysis models are different, but both are compatible with a common bivariate normal model of $(X,Y)$. It follows that matrix calculations should involve all 5 parameters of the bivariate normal model (see section \ref{section:Design}). To see why, let's examine what happens when matrix calculations are limited to just two parameters: the slope and intercept of the analysis model. 

\begin{table}
\caption{Estimating the slope of X on Y. Mean length of 95\% confidence intervals (CIs). \\ (Parentheses enclose \% departure from 95\% coverage.)}
\label{tab:X_on_Y}

\begin{subtable}{\textwidth}
            \centering
            \caption{With matrix formulas limited to two parameters ($\alpha_{X.Y},\beta_{X.Y}$).}
            \label{tab:X_on_Y_2}
\begin{tabular}{ccccccccc} 
\hline
\multicolumn{2}{c}{Missing} &             & \multicolumn{2}{c}{Score-based CIs}    & \multicolumn{2}{c}{Within-between CIs}    & \multicolumn{2}{c}{Bootstrapped CIs}     \\ 
\cline{1-2}\cline{4-9}
\% & Pattern                & Imputations & PDMI        & MLMI        & PDMI        & MLMI        & PDMI         & MLMI         \\ 
\hline
25 & MCAR                   & 10          & 0.17 (0.6)  & 0.17 (0.5)  & 0.17 (0.3)  & 0.17 (0.9)  &              &              \\
   &                        & 50          & 0.17 (0.7)  & 0.17 (0.6)  & 0.17 (0.5)  & 0.17 (0.3)  & 0.18 (0.1)   & 0.18 (0.5)   \\
   &                        & 200         & 0.17 (0.3)  & 0.17 (0.2)  & 0.17 (0.0)  & 0.17 (-0.2) & 0.17 (-0.2)  & 0.17 (-0.1)  \\
   & MAR                    & 10          & 0.17 (-0.2) & 0.17 (-0.1) & 0.17 (-0.3) & 0.17 (0.1)  &              &              \\
   &                        & 50          & 0.17 (0.1)  & 0.17 (0.0)  & 0.17 (-0.1) & 0.17 (-0.1) & 0.18 (-0.2)~ & 0.18 (-0.4)  \\
   &                        & 200         & 0.17 (0.1)  & 0.17 (0.0)  & 0.17 (0.1)  & 0.17 (-0.4) & 0.17 (-0.3)  & 0.17 (-0.2)  \\
50 & MCAR                   & 10          & 0.20 (0.2)  & 0.20 (0.0)  & 0.20 (0.0)  & 0.20 (-0.2) &              &              \\
   &                        & 50          & 0.20 (0.1)  & 0.20 (0.0)  & 0.20 (0.1)  & 0.19 (-1.1) & 0.21 (-0.2)  & 0.21 (-0.4)  \\
   &                        & 200         & 0.20 (0.5)  & 0.20 (0.5)  & 0.20 (0.4)  & 0.19 (-0.9) & 0.20 (0.3)   & 0.20 (-0.1)  \\
   & MAR                    & 10          & 0.21 (-1.4) & 0.20 (-1.2) & 0.22 (0.0)  & 0.20 (-1.7) &              &              \\
   &                        & 50          & 0.20 (-1.3) & 0.20 (-1.5) & 0.21 (0.3)  & 0.19 (-2.8) & 0.23 (0.3)   & 0.23 (0.0)   \\
   &                        & 200         & 0.20 (-1.0) & 0.20 (-1.2) & 0.21 (0.5)  & 0.19 (-2.9) & 0.21 (0.2)   & 0.21 (0.1)   \\
\hline
\end{tabular}
\end{subtable}

\begin{subtable}{\textwidth}
            \centering
            \caption{With matrix formulas including all five parameters ($\alpha_{X.Y},\beta_{X.Y},\sigma^2_{X.Y},\mu_Y,\sigma^2_Y,$).}
\begin{tabular}{ccccccc} 
\hline
\multicolumn{2}{c}{Missing} &             & \multicolumn{2}{c}{Score-based CIs}  & \multicolumn{2}{c}{Within-between CIs}    \\ 
\cline{1-2}\cline{4-7}
\% & Pattern                & Imputations & PDMI       & MLMI       & PDMI       & MLMI         \\ 
\hline
25 & MCAR                   & 10          & 0.17 (0.8) & 0.17 (0.7) & 0.17 (0.3)  & 0.18 (1.2)   \\
   &                        & 50          & 0.17 (0.8) & 0.17 (0.9) & 0.17 (0.5)  & 0.17 (0.5)   \\
   &                        & 200         & 0.17 (0.5) & 0.17 (0.4) & 0.17 (0.0)  & 0.17 (0.1)   \\
   & MAR                    & 10          & 0.17 (0.1) & 0.17 (0.2) & 0.17 (-0.3) & 0.18 (0.4)   \\
   &                        & 50          & 0.17 (0.3) & 0.17 (0.3) & 0.17 (-0.1) & 0.17 (0.3)   \\
   &                        & 200         & 0.17 (0.3) & 0.17 (0.3) & 0.17 (0.1)  & 0.17 (0.0)   \\
50 & MCAR                   & 10          & 0.21 (0.5) & 0.20 (0.3) & 0.20 (0.0)  & 0.21 (0.7)   \\
   &                        & 50          & 0.20 (0.4) & 0.20 (0.2) & 0.20 (0.1)  & 0.21 (0.4)   \\
   &                        & 200         & 0.20 (0.8) & 0.20 (0.8) & 0.20 (0.4)  & 0.20 (0.5)   \\
   & MAR                    & 10          & 0.22 (0.8) & 0.22 (0.6) & 0.22 (0.0)  & 0.21 (-0.9)  \\
   &                        & 50          & 0.22 (0.6) & 0.22 (0.5) & 0.21 (0.3)  & 0.22 (0.2)   \\
   &                        & 200         & 0.22 (0.9) & 0.21 (0.5) & 0.21 (0.5)  & 0.22 (0.5)   \\
\hline
\end{tabular}
\end{subtable}
\end{table}

Table \ref{tab:X_on_Y_2} summarizes CIs for the slope $\beta_{X.Y}$. The bootstrap and PDMI WB CIs have good coverage under all simulated conditions, but the other CIs do not. Under most simulated conditions, all CIs have good coverage, but when 50\% of values are MAR, the WB intervals undercover under MLMI, and the SB intervals undercover under both PDMI and MLMI. This undercoverage does not improve as the number of imputations increases.

The reason for the undercoverage is that the SB and MLMI  WB formulas have underestimated the covariance matrix of the estimates. That's because we limited the SB and PDMI WB formulas to the $2\times2$ covariance matrices associated with the two parameters ($\alpha_{X.Y},\beta_{X.Y}$).

But consistent estimation requires that we apply the SB and MLMI WB formulas to the full $5\times5$ matrix describing the five parameters of the bivariate normal model $(\mu_Y,\sigma^2_Y,\alpha_{X.Y},\beta_{X.Y},\sigma^2_{X.Y})$ . 
 
Table \ref{tab:X_on_Y}B shows what happens when we do that. The covariance matrices are now consistently estimated, and the confidence intervals have close to nominal coverage.

Although the need to use all five parameters in variance calculations is somewhat limiting, in the simulation it only made a noticeable difference when the fraction of missing information was quite large (i.e., 50\% of values MAR). When the fraction of missing information was small to moderate, as it often is in applied work, neglecting parts of the parameter vector yielded acceptable results. In the next section, we will also get acceptable estimates when applying these methods to an applied dataset.

\section{Applied data analysis}
We next conducted an applied data analysis to compare MLMI to PDMI with different approaches to variance estimation. We analyzed data from the Millennium Cohort Study \cite{mcsData}, a longitudinal cohort study that followed approximately 19,000 children who were born between 2000 and 2001 in the United Kingdom. We analyzed data from wave 2 of the study, when the children were around 3 years old. 

Our imputation model was a general location model, which consisted of a log-linear model of the categorical variables and a conditionally multivariate normal model of the continuous variables \cite{Schafer1997}. The log-linear model included all 2-way interactions, and the mean of the multivariate normal distribution depended only on main effects of the categorical variables. The imputation model used two auxiliary variables, which were not in the analysis model but improved the imputation of variables that were \cite{vHLynch2013}. One auxiliary variable was the marital status of the parents; the other was the employment status of the parent or guardian responding to the survey. 
 
We multiply imputed missing values using both MLMI and PDMI. Under MLMI, we obtained ML parameter estimates using the EM algorithm. Under PDMI, we obtained PD parameter estimates with an MCMC algorithm that started with 100 burn-in iterations and then drew every 100th estimate from the Markov chain. 

Our analysis model was a linear regression of each child's school readiness, as measured by the Bracken score, on family income, tenure of housing, any history of the child having hearing loss, ethnicity, number of siblings (categorized as 1, 2, 3+), and the age of the parent or guardian responding to the survey. The percentage of missing values varied from 0.013\% for the number of siblings to 15.8\% for family income. The Millennium Cohort Study uses a complex sampling scheme, but for simplicity of illustration we analyzed it as though it were a simple random sample.

For our first analysis we used 100 imputations; for our second, we used 1,000 to approximate the asymptotic behavior of the estimators. When we used repeated MI, the number of imputations was $M$; when we used bootstrapped MI, the number of imputations was $BD$, where $B=50$ (in the first analysis) or 500 (in the second) was the number of bootstrapped samples, and $D=2$ was the number of imputations per bootstrapped sample. We analyzed the imputed data using the linear regression model described above, applying WB, SB, and bootstrap formulas to get SEs for the parameters of the analysis model. 

\subsection{Results with 100 imputations}
Table \ref{tab:MCS_100} shows results with 100 imputations. Table \ref{tab:MCS_100_runtimes} gives the runtime (in seconds) needed to impute the data 100 times and analyze it on a personal computer (a Dell Latitude 7400 with an i7 CPU and 16GB RAM). Although all runtimes were under a minute, imputing was much faster with MLMI than with PDMI. When we used repeated imputation, MLMI was 25 times faster than PDMI; when we used bootstrapped imputation, MLMI was 4 times faster than PDMI. Bootstrapped MLMI, though 9 times slower than repeated MLMI, was still 3 times faster than repeated PDMI. After imputation, the calculation of SEs took approximately the same runtime under MLMI as under PDMI. Score-based SE formulas were 3 times slower than other SE formulas.

The slowness of PDMI was due in part to the iterative MCMC algorithm that implemented it \cite{Schafer1997}. While MCMC is the most common PDMI algorithm, the bootstrapped EM algorithm makes PDMI faster \cite{King2001}, though still not as fast as MLMI.

Table \ref{tab:MCS_100_point_estimates} compares point estimates of the regression parameters. Among the MI estimates, the MLMI and PDMI estimates are very similar, with or without the bootstrap. This empirical result is consistent with our theoretical results showing that MI point estimates, with or without the bootstrap, are close to their asymptotic values when 100 imputations are used. The MI point estimates differ by less than 10 percent complete case estimates, except for the coefficient of ``Other housing,'' which differs by a factor of 4.

Table \ref{tab:MCS_100_SEs} compares SE estimates for the regression parameters. Under repeated MI, nearly the SE estimates are very similar whether we use MLMI or PDMI, and whether we used score-based or within-between formulas.\footnote{The one discrepancy is the SE of the ``non-white'' coefficient, which is about 15 percent larger using the within-between formula than using the score-based formula.} This empirical result is consistent with our theoretical results showing that, with 100 imputations, score-based and within-between variance formulas come close to their asymptotic values.

Under bootstrapped MI, many of the SE estimates are similar under MLMI and PDMI, but there are a few noticeable differences. This reflects the fact that bootstrapped SE estimates can be somewhat variable when there are only $B=50$ bootstrapped samples. With $B=50$, the coefficient of variation for a bootstrapped SE estimates is about 10\%\footnote{As discussed earlier, the coefficient of variation for an SE estimate is approximately $\sqrt{1/(2df)}$, and under bootstrapped MI $df$ is just a little smaller than $B$.}, implying that a bootstrapped SE estimate typically changes by about 10\% when the data are bootstrapped and imputed again. That explains most of the differences between the bootstrapped SE estimates obtained under MLMI and PDMI. The differences do not reflect a difference between MLMI and PDMI; we would see similar differences if we had used bootstrapped MLMI twice, or bootstrapped PDMI twice. When $B$ is larger, bootstrapped SE estimates are less variable and agree more closely under MLMI and PDMI --- as we will show next.

\begin{table}
\caption{Imputation and analysis of Millennium Cohort Study, using 100 imputations ($M=100$ under repeated MI, $B=50, D=2$ under bootstrapped MI)}
\label{tab:MCS_100}

\begin{subtable}{\textwidth}
\caption{Runtime (in seconds)}
\label{tab:MCS_100_runtimes}
\begin{tabular}{lccc}
  \hline
 & MLMI & PDMI & Runtime ratio (PDMI/MLMI) \\ 
  \hline
  Repeated imputation & 1.5 & 37.2 & 25 \\ 
  Bootstrapped imputation & 13.3 & 52.6 & 4 \\ 
  Within-between SE calculation & 8.7 & 7.6 & 1 \\ 
  Score-based SE calculation & 21.5 & 20.6 & 1\\ 
  Bootstrap SE calculation & 7.3 & 7.1 & 1 \\ 
  \hline
\end{tabular}
\end{subtable}

\begin{subtable} {\textwidth}
            \caption{Regression point estimates}
            \label{tab:MCS_100_point_estimates}
\begin{tabular}{lccccc}
  \hline
  &\multicolumn{2}{c}{Repeated MI}&\multicolumn{2}{c}{Bootstrapped MI} \\
 \cline{2-3} \cline{4-5} 
 & MLMI & PDMI & MLMI & PDMI & Complete case analysis \\
  \hline
Intercept & 89.78 & 89.84 & 89.70 & 89.85 & 89.12 \\ 
  Parent or guardian's age (years) & 0.24 & 0.24 & 0.23 & 0.24 & 0.27 \\ 
  Family income & 0.90 & 0.90 & 0.90 & 0.90 & 0.87 \\ 
  Rented housing & -3.83 & -3.88 & -3.89 & -3.89 & -4.03 \\ 
  Other housing & -0.47 & -0.53 & -0.59 & -0.45 & -0.14 \\ 
  Child hearing loss & 2.74 & 2.72 & 2.84 & 2.67 & 3.02 \\ 
  Non-white & -7.08 & -7.06 & -7.08 & -7.11 & -6.40 \\ 
  1 sibling & -2.43 & -2.45 & -2.42 & -2.46 & -2.52 \\ 
  2 siblings & -6.73 & -6.73 & -6.67 & -6.80 & -6.70 \\ 
  3 or more siblings & -10.77 & -10.74 & -10.77 & -10.79 & -10.59 \\ 
   \hline
\end{tabular}
\end{subtable}

\begin{subtable}{\textwidth}
\caption{Regression standard error estimates}
 \label{tab:MCS_100_SEs}
\begin{tabular}{lcccccc}
  \hline
  &\multicolumn{4}{c}{Repeated MI}&& \\
   \cline{2-5} 
    &\multicolumn{2}{c}{Score-based SEs}&\multicolumn{2}{c}{Within-between SEs}&\multicolumn{2}{c}{Bootstrapped SEs} \\
 \cline{2-3} \cline{4-5} \cline{6-7} 

 & MLMI  & PDMI  & MLMI  & PDMI  & MLMI  & PDMI  \\ 
  \hline
Intercept & 1.04 & 1.04 & 1.06 & 1.04 & 1.09 & 1.11 \\ 
  Parent or guardian's age (years) & 0.02 & 0.02 & 0.02 & 0.02 & 0.03 & 0.03 \\ 
  Family income & 0.04 & 0.04 & 0.05 & 0.04 & 0.04 & 0.05 \\ 
  Rented housing & 0.32 & 0.32 & 0.34 & 0.33 & 0.35 & 0.35 \\ 
  Other housing & 0.71 & 0.71 & 0.72 & 0.73 & 0.69 & 0.75 \\ 
  Child hearing loss & 0.65 & 0.65 & 0.62 & 0.62 & 0.64 & 0.57 \\ 
  Non-white & 0.36 & 0.36 & 0.41 & 0.42 & 0.47 & 0.38 \\ 
  1 sibling & 0.31 & 0.31 & 0.31 & 0.31 & 0.35 & 0.34 \\ 
  2 siblings & 0.38 & 0.38 & 0.39 & 0.38 & 0.45 & 0.46 \\ 
  3 or more siblings & 0.48 & 0.48 & 0.48 & 0.49 & 0.48 & 0.58 \\
   \hline
\end{tabular}
\end{subtable}
\end{table}

\subsection{Results with 1,000 imputations}
Table \ref{tab:MCS_1000} shows results for 1,000 imputations. Table \ref{tab:MCS_1000_runtimes} compares runtimes. With 1,000 imputations, MLMI was still much faster than PDMI. Under repeated imputation, MLMI took half a minute, while PDMI took six and a half minutes. Under bootstrapped imputation, MLMI took two minutes, while PDMI took eight and a half. MLMI's runtime advantage of approximately six minutes was substantial, and could affect analysts' productivity and morale, especially if they re-specified the imputation model and re-imputed the data several times.

Table \ref{tab:MCS_1000_point_estimates} compares regression point estimates. The estimates are very similar under MLMI and PDMI, with or without the bootstrap. In fact, the point estimates with 1,000 imputations are very close to the point estimates that we obtained with 100 imputations (Table \ref{tab:MCS_100_point_estimates}), confirming our claim that those point estimates were close to their asymptotic values. 

Table \ref{tab:MCS_1000_SEs} compares SE estimates. With 1,000 imputations, nearly all the SE estimates are very similar, whether we used MLMI or PDMI with the bootstrap, the score-based formula, or the within-between formula. Evidently 1,000 imputations was enough to bring all the SE estimates close to their asymptotic values. The bootstrapped SE estimates were the most variable, but with $B=500$ they typically came within 3 percent of their asymptotic values.\footnote{As discussed earlier, the coefficient of variation for an SE estimate is approximately $\sqrt{1/(2df)}$, and under bootstrapped MI $df$ is just a little smaller than $B$. So with $B=500$, the coefficient of variation for a bootstrapped SE estimate is 3 percent.}

When there are substantial disagreements between different SE estimates, we favor the bootstrapped estimates because $B$ is large and the bootstrap is consistent even when the imputation and analysis models are incompatible or misspecified. For example, for the coefficient non-white children, the true SE is probably closer to the 0.43 given by the bootstrap than to the 0.36 given by the SB formulas or the 0.40-0.41 given by the WB formulas. But such disagreements are rare.

How surprised should we be that the different SE formulas agree so well? There are two considerations. First, the formulas make different assumptions about the imputation and analysis models (section \ref{section:assumptions}). 
\begin{itemize}
\item The bootstrap SE formulas are consistent even when the imputation and analysis models are incompatible or misspecified. So they are consistent here. 
\item The PDMI WB formula is consistent when the imputation and analysis models are compatible and correct. Here the imputation and analysis models are compatible \cite{Bartlett2015}, and although they are unlikely to be perfectly specified, evidently any misspecification is not serious enough to introduce much bias. If there were much bias, we would more often see the PDMI WB SEs disagreeing with the bootstrap.
\item The SB and MLMI WB matrix formulas have additional requirements. Not only must the imputation and analysis models be consistent and correct, but the SB and MLMI WB matrices should include parameters from the imputation model that are not in the analysis model. In this example, though, the matrices included only  parameters from the analysis model --- and returned SE estimates that were mostly similar to the consistent bootstrapped estimates.
\end{itemize}

Perhaps a reason for the near-agreement across different formulas is that the fraction of missing information is rather small. In our simulations, we found that the differences among SE estimates were barely noticeable unless the fraction of missing information was quite large. 

\begin{table}
\caption{Imputation and analysis of Millennium Cohort Study, using 1,000 imputations ($M=1,000$ under repeated MI, $B=500, D=2$ under bootstrapped MI)}
\label{tab:MCS_1000}
\begin{subtable}{\textwidth}
\caption{Runtime (in seconds)}
\label{tab:MCS_1000_runtimes}
\begin{tabular}{lccc}
  \hline
 & MLMI & PDMI & Runtime ratio (PDMI/MLMI) \\ 
  \hline
  Repeated imputation & 33.4 & 394.5 & 12 \\ 
  Bootstrapped imputation & 123.6 & 512.4 & 4 \\ 
  Within-between SE calculation & 78.2 & 75.2 & 1 \\ 
  Score-based SE calculation & 217.0 & 214.5 & 1 \\ 
  Bootstrap SE calculation & 64.5 & 65.9 & 1 \\ 
   \hline
\end{tabular}
\end{subtable}

\begin{subtable} {\textwidth}
            \caption{Regression point estimates}
            \label{tab:MCS_1000_point_estimates}
\begin{tabular}{lccccc}
  \hline
  &\multicolumn{2}{c}{Repeated MI}&\multicolumn{2}{c}{Bootstrapped MI} \\
 \cline{2-3} \cline{4-5} 
 & MLMI & PDMI & MLMI & PDMI & Complete case analysis \\
  \hline
Intercept & 89.81 & 89.84 & 89.81 & 89.80 & 89.12 \\ 
  Parent or guardian's age (years) & 0.24 & 0.24 & 0.24 & 0.24 & 0.27 \\ 
  Family income & 0.90 & 0.89 & 0.89 & 0.90 & 0.87 \\ 
  Rented housing & -3.85 & -3.86 & -3.86 & -3.86 & -4.03 \\ 
  Other housing & -0.50 & -0.50 & -0.51 & -0.59 & -0.14 \\ 
  Child hearing loss & 2.74 & 2.72 & 2.74 & 2.71 & 3.02 \\ 
  Non-white & -7.08 & -7.07 & -7.08 & -7.08 & -6.40 \\ 
  1 sibling & -2.44 & -2.44 & -2.44 & -2.43 & -2.52 \\ 
  2 siblings & -6.74 & -6.75 & -6.72 & -6.73 & -6.70 \\ 
  3 or more siblings & -10.78 & -10.78 & -10.78 & -10.78 & -10.59 \\ 
   \hline
\end{tabular}
\end{subtable}

\begin{subtable}{\textwidth}
\caption{Regression standard error estimates}
 \label{tab:MCS_1000_SEs}
\begin{tabular}{lcccccc}
  \hline
    &\multicolumn{2}{c}{Score-based SEs}&\multicolumn{2}{c}{Within-between SEs}&\multicolumn{2}{c}{Bootstrapped SEs} \\
 \cline{2-3} \cline{4-5} \cline{6-7} 

 & MLMI  & PDMI  & MLMI  & PDMI  & MLMI  & PDMI  \\ 
  \hline
Intercept & 1.04 & 1.04 & 1.03 & 1.03 & 1.02 & 1.06 \\ 
  Parent or guardian's age (years) & 0.02 & 0.02 & 0.02 & 0.02 & 0.03 & 0.03 \\ 
  Family income & 0.04 & 0.04 & 0.04 & 0.04 & 0.04 & 0.04 \\ 
  Rented housing & 0.32 & 0.32 & 0.33 & 0.33 & 0.35 & 0.35 \\ 
  Other housing & 0.71 & 0.71 & 0.72 & 0.72 & 0.67 & 0.73 \\ 
  Child hearing loss & 0.65 & 0.65 & 0.62 & 0.62 & 0.60 & 0.62 \\ 
  Non-white & 0.36 & 0.36 & 0.41 & 0.40 & 0.43 & 0.43 \\ 
  1 sibling & 0.31 & 0.31 & 0.31 & 0.31 & 0.32 & 0.33 \\ 
  2 siblings & 0.38 & 0.38 & 0.38 & 0.39 & 0.39 & 0.38 \\ 
  3 or more siblings & 0.48 & 0.48 & 0.48 & 0.48 & 0.48 & 0.48 \\  
   \hline
\end{tabular}
\end{subtable}
\end{table}

\section{Conclusion}
MLMI offers a serious alternative to PDMI. MLMI is not the only alternative --- fractional imputation also deserves serious consideration \cite{Yang2016} --- but it does have certain advantages over PDMI. 

The first advantage of MLMI is its computational efficiency. MLMI is easier to code than PDMI, and MLMI runs faster: it can produce more imputations in the same runtime. The speed advantage of MLMI is substantial when PDMI uses MCMC to get posterior draws, as most PDMI software does. The speed advantage of MLMI is more modest when PDMI gets posterior draws with a more efficient algorithm, such as bootstrapped ML \cite{King2001}. 

The second advantage of MLMI is the efficiency of its point estimates. Compared to PDMI point estimates, MLMI point estimates are more efficient when they use the same number of imputations as PDMI, and still more efficient when MLMI uses the larger number of imputations that it can generate in the same runtime as PDMI. The efficiency advantage of MLMI point estimates is typically quite small, but can be larger when the fraction of missing information is large and PDMI uses few imputations.  

Until now, the use of MLMI has been discouraged by the lack of convenient formulas for variances, SEs, and CIs. But we have derived and evaluated three SE estimators: the within-between (WB) estimator, the score-based (SB) estimator, and bootstrapped MI. Some of these SE estimators are more viable than others.

The WB variance formulas use variance components that lie within and between the imputed datasets. An old WB formula \eqref{V_PDMI_WB} has been used with PDMI for over 30 years \cite{Rubin1987}, and we have derived a new WB formula \eqref{V_MLMI_WB} that is consistent under MLMI. Our MLMI WB formula requires more imputations than the PDMI WB imputations, but when the fraction of missing information is 50\% or less, the number of imputations required is not excessive and often present no practical problem since MLMI produces imputations more quickly than PDMI (Table \ref{tab:imputations_for_df}). With more than 50\% missing information, though, the MLMI WB formula requires a rapidly increasing number of imputations, so that it becomes better to use bootstrapped MI, which with high missing information can produce better SE estimates with fewer imputations. 

The SB variance formulas decompose the variance of the score function. The same SB formula is consistent under PDMI and MLMI. The SB variance formula needs fewer imputations than the WB formulas, but its calculation requires the contribution of each case to the score function. This can be a serious disadvantage, since the user often does not know the contribution of each case to the score function, and some approaches to estimation do not use the score function at all. This limits the practical use of the SB formula.

When the imputation and analysis models are the same, both the SB formula and the MLMI WB formula can be applied to the parameters of the analysis model alone. But when the imputation and analysis models are different, the SB and MLMI WB formulas can also require the parameters of the underlying common model that is consistent with both the imputation and analysis model. When these additional parameters are neglected, the SB and MLMI WB formulas can produce poor SE estimates, although in practice the SE estimates seem to perform well unless the fraction of information is quite large. 

Bootstrapped MI variance estimation is the most robust approach. It is flexible and can work with a variety of imputation methods, including but not limited to PDMI and MLMI. Bootstrapped MI variance estimates are consistent even when the imputation and analysis models are different or misspecified. Unlike SB and MLMI WB estimates, bootstrapped MI estimates never require parameter estimates beyond those from the analysis model. Unlike WB variance estimates, bootstrapped MI variance estimates do not require a complete-data analytic SE for complete data, and so can be used in situations where analytic SEs are unavailable or invalid. 

A further advantage of bootstrapped MI variance estimates is that they are consistent even when the imputation and analysis models are incompatible or misspecified. This property is valuable since in practical settings most models are at least a little misspecified, and incompatibility between the imputation and analysis models is common. While no method can ensure that \textit{point estimates} will be consistent under a misspecified model, bootstrapped MI can at least ensure that the variability of point estimates is estimated accurately. This is a property that WB and SB estimates lack, under both MLMI and PDMI. We know of only one other approach that can produce consistent variance estimates under misspecified and incompatible imputation and analysis models \cite{Robins2000} --- but the calculations are relatively complicated and require statistics, including but not limited to the score function, that users often lack access to in practical settings. 

Bootstrapped MI, by contrast, is straightforward. An old knock against bootstrapped MI was that it seemed to require a large number of imputations $D$ for each bootstrapped sample \cite{Rubin1994}. Our approach, however, produces consistent variance estimates with just $D=2$ imputations. Another knock was that the bootstrap can require a large number of bootstrapped samples $B$, but that requirement is not limited to imputed data. In complete data, the bootstrap can also require a large $B$, and analysts often consider that an acceptable price to pay for robust SE estimates. Bootstrapping MI requires approximately the same $B$ as bootstrapping complete data. In both complete and MI data, the degrees of freedom is slightly less than $B$, and perhaps $B=25$ samples suffice for replicable point estimates, and $B=500$ for replicable SE estimates. Imputing $B$ bootstrapped samples can take a long time if you use PDMI, but MLMI can impute the bootstrapped samples much more quickly.

\clearpage
\appendix

\section{ Simplified expression for ${\boldsymbol{V}}_{\boldsymbol{PDMI}}$}
\label{appendix: simplify_V_PDMI}

In equation \eqref{V_PDMI} we gave an expression for $V_{PDMI}$ which we claimed was equivalent to the more complicated expression in equation (2) of Wang and Robins (1998). Below we give the steps of the simplification. The first line gives Wang and Robins' equation (2), with a typo corrected and the symbols changed to match our notation. The last line gives our simplified expression \eqref{V_PDMI}.
\begin{eqnarray*}
V_{PDMI} &=& V_{ML}+\frac1{M}V_{com}{\gamma }_{mis}+\frac1{M}{\gamma }^T_{mis}V_{ML}{\gamma }_{mis} \\ 
&=& V_{ML}+\frac1{M}(V_{com}+{\gamma }^T_{mis}V_{ML}){\gamma }_{mis}\\ &=& V_{ML}+\frac1{M}(V_{com}+{(I-{\gamma }_{obs})}^TV_{ML}){\gamma }_{mis} \\ 
&=& V_{ML}+\frac1{M}{(V}_{com}+{(I-V^{-1}_{ML}V_{com})}^TV_{ML}){\gamma }_{mis} \\ 
&=& V_{ML}+\frac1{M}{(V}_{com}+{((V^{-1}_{ML}(V_{ML}-V_{com}))}^TV_{ML}){\gamma }_{mis} \\ 
&=& V_{ML}+\frac1{M}{(V}_{com}+{(V_{ML}-V_{com})}^TV^{-T}_{ML}V_{ML}){\gamma }_{mis} \\ 
&=& V_{ML}+\frac1{M}{(V}_{com}+(V_{ML}-V_{com})V^{-1}_{ML}V_{ML}){\gamma }_{mis} \\ 
&=& V_{ML}+\frac1{M}(V_{com}+V_{ML}-V_{com}){\gamma }_{mis} \\ 
&=& V_{ML}+\frac1{M}V_{ML}{\gamma }_{mis} 
\end{eqnarray*}

\section{Shrinking WB estimates under MLMI}
\label{appendix:shrinkage_estimator}

In section \ref{subsection:MLMI_WB} we presented a simple estimator $\widehat{\gamma }_{mis|MLMI,WB}=\widehat{W}^{-1}_{MLMI}{\widehat{B}}_{MLMI}$ for the fraction of missing information under MLMI, then replaced it with the shrunken estimator ${\widetilde{\gamma }}_{mis \mid MLMI}=h(\widehat{\gamma },M-1)$. We now explain why shrinkage is necessary, and justify our shrinkage function $h()$.

\begin{figure}
\caption{The probability that $\widehat{\gamma }_{mis \mid MLMI}$ exceeds 1, as a function of $m$ and ${\gamma }_{mis}$.}
\label{fig:prob_gamma_1}
\includegraphics[width=\textwidth]{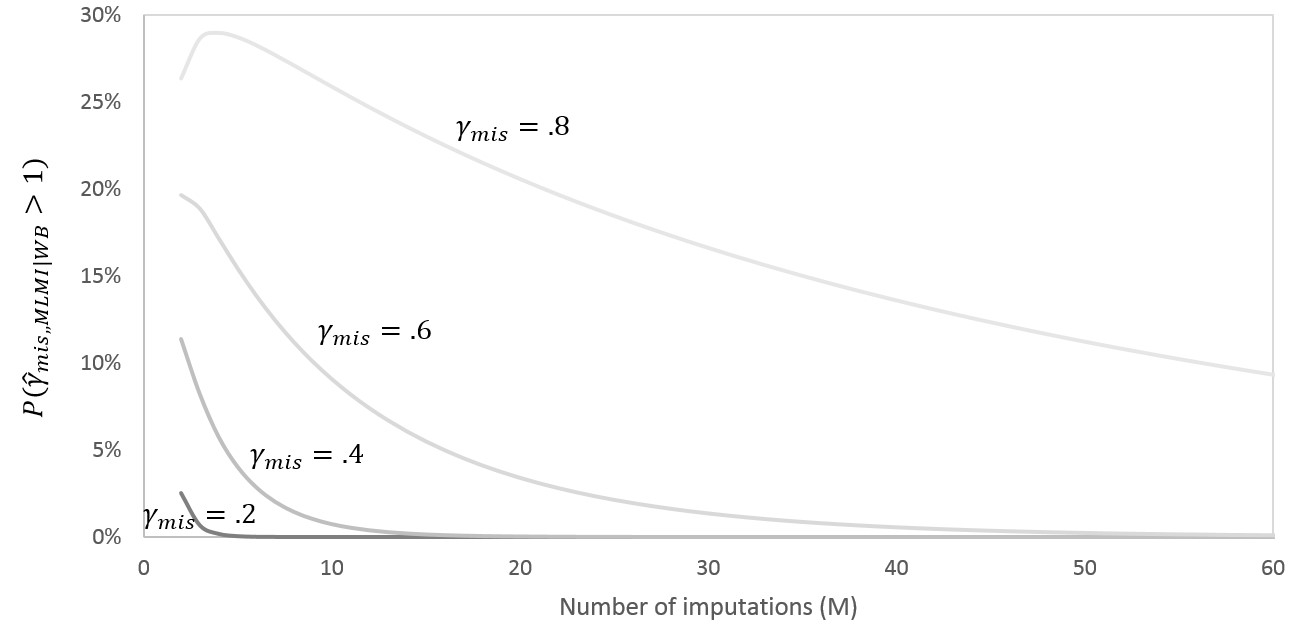}
\end{figure}

The problem with the simple estimator $\widehat{\gamma }_{mis|MLMI,WB}$ is that it can exceed 1, whereas the true fraction of missing information ${\gamma }_{mis}$ cannot. To show this, we adopt the convention, common in the MI literature, that the variation in ${\widehat{W}}_{MI}$ is negligible compared to the variation in ${\widehat{B}}_{MI}$. Then the distribution of ${\widehat{\gamma }}_{mis \mid MLMI}$ is approximately scaled chi-square:

\begin{equation} 
\widehat{\gamma }_{mis \mid MLMI} = {\gamma }_{mis}\frac{U}{M-1}\mathrm{,\ where\ }U\mathrm{\sim }{\chi }^{2}_{M-1}
\end{equation}

\noindent and the probability that ${\widehat{\gamma }}_{mis \mid MLMI}$ exceeds 1 is $P\left({\gamma }_{mis}\frac{U}{M\mathrm{-}\mathrm{1}}>1\right)=P\left(U>\frac{M\mathrm{-}\mathrm{1}}{{\gamma }_{mis}}\right)$. Figure \ref{fig:prob_gamma_1} graphs this probability as a function of $m$ and ${\gamma }_{mis}$. The probability is negligible if ${\gamma }_{mis}$ is low, but can be substantial if ${\gamma }_{mis}$ is high and $M$ is low relative to ${\gamma }_{mis}$.

Our solution is to replace ${\widehat{\gamma }}_{mis\mathrm{|}MLMI,WB}$ with a shrunken estimator ${\widetilde{\gamma }}_{mis\mathrm{|}MLMI,WB}$ which is guaranteed to take values in (0,1). We define ${\widetilde{\gamma }}_{mis\mathrm{|}MLMI,WB}$ as the posterior mean of ${\gamma }_{mis}$ when the prior is uniform on (0,1). With this prior, the posterior distribution of ${\gamma }_{mis}$ approximates a scaled inverse chi-square---
\begin{equation}
{\gamma }_{mis}=\widehat{\gamma }_{mis,|MLMI,WB}\frac{M-1}{U}, \text{where } U\sim{\chi }^{2}_{M-1}
\end{equation}

\noindent ---with the modification that the distribution of ${\gamma }_{mis}$ is truncated on the right at 1. We calculated the mean of this truncated distribution using Mathematica software, version 8. The solution is \eqref{shrink_scalar}---i.e., 
\begin{equation} 
{\widetilde{\gamma }}_{mis|MLMI,WB} = h({\widehat{\gamma }}_{mis|MLMI,WB},M-1)
\end{equation}

\noindent where
\begin{equation}
h(\widehat{\gamma },\nu)=\frac{\nu }{2}\widehat{\gamma }\frac{{\mathit{\Gamma} \left(\frac{\nu -2}{2},\frac{\nu }{2}\widehat{\gamma }\right)\ }}{{\mathit{\Gamma} \left(\frac{\nu }{2},\frac{\nu }{2}\widehat{\gamma }\right)\ }}
\end{equation}

Using numerical integration in Mathematica software, we calculate the bias $E(\widetilde{\gamma }_{mis|MLMI,WB}-{\gamma}_{mis})$ that is summarized in Table \ref{tab:imputations_for_bias}.

Since the function $\mathit{\Gamma}(a,z)$ is unavailable in some statistical software, for implementation purposes it helps to know that with $\nu >2$, $h(\widehat{\gamma },\nu)$ simplifies to
\begin{equation} 
h(\widehat{\gamma },\nu)=\frac{\nu }{\nu -2}\widehat{\gamma }\frac{R_{\mathrm{\Gamma }}\left(\frac{\nu -2}{2},\frac{\nu }{2}\widehat{\gamma }\right)}{R_{\mathrm{\Gamma }}\left(\frac{\nu }{2},\frac{\nu }{2}\widehat{\gamma }\right)}
\end{equation}

\noindent where $R_{\mathrm{\Gamma }}(a,z)$, which is widely available in statistical software, is the survival function for a gamma distribution with shape parameter $a$, evaluated at $z$. Since this simplification requires $\nu >2$, it can only be used when $M>$4.

\section{ Degrees of freedom for WB variance estimation under MLMI}
\label{appendix:df}

Equation \eqref{df_MLMI_WB} approximates the $df$  of the variance estimate $\widetilde{V}_{MLMI,WB}$. Although $\widetilde{V}_{MLMI,WB}$ is not a chi-square variable, a chi-squared variable with $df={\widehat{\nu }}_{MLMI,WB}$ will have approximately the same coefficient of variation (CV) as ${\widetilde{V}}_{MLMI,WB}$.

To derive this approximation, consider the scalar expression 
\begin{equation}
\widetilde{V}_{MLMI,WB}={\widetilde{V}}_{ML|MLMI,WB}+\frac1{M}{\widehat{B}}_{MLMI}
\end{equation}

where
\begin{eqnarray}
\widetilde{V}_{ML|MLMI,WB} &=& {\widehat{W}}_{MLMI}{\widetilde{\gamma }}^{-1}_{obs\mathrm{|}MLMI,WB}  \\{\widetilde{\gamma }}_{obs\mathrm{|}MLMI,WB}
 &=& 1-{\widetilde{\gamma }}_{mis\mathrm{|}MLMI,WB}  \\ {\widetilde{\gamma }}_{mis\mathrm{|}MLMI,WB} 
 &=& h\left(\mathrm{\ }{\widehat{\gamma }}_{mis\mathrm{|}MLMI,WB}\right) \\{\widehat{\gamma }}_{mis\mathrm{|}MLMI,WB} 
 &=& {\widehat{W}}^{-1}_{MLMI}{\widehat{B}}_{MLMI} 
\end{eqnarray}

We can approximate the distribution of ${\widetilde{V}}_{MLMI,WB}$ by starting with its components. ${\widehat{B}}_{MLMI}$ has approximately a scaled ${\chi }^2_{M-1}$ distribution, and if we regard ${\widehat{W}}_{MLMI}$ as fixed, then ${\widehat{\gamma }}_{mis\mathrm{|}MLMI,WB}$ also has approximately a scaled ${\chi }^2_{M-1}$ distribution with expectation ${\gamma }_{mis}$. We regard ${\widetilde{\gamma }}_{mis\mathrm{|}MLMI,WB}$ as having approximately the same distribution as ${\widehat{\gamma }}_{mis\mathrm{|}MLMI,WB}$.

Under these assumptions, ${\widetilde{\gamma }}_{obs\mathrm{|}MLMI,WB}$ has expectation ${\gamma }_{obs}$, standard deviation ${\gamma }_{mis}\sqrt{2/\ (M-1)}$, and CV=$\left(\frac{{\gamma }_{mis}}{{\gamma }_{obs}}\right)\sqrt{2/(M-1)}$, which is also the CV of a ${\chi }^2_{{\nu }_1}$ variable with $df={\nu }_1=(M-1){\left(\frac{{\gamma }_{obs}}{{\gamma }_{mis}}\right)}^2$. So we can approximate ${\widetilde{\gamma }}_{obs\mathrm{|}MLMI,WB}$ as a scaled ${\chi }^2_{{\nu }_1}$ variable.

Then ${\widetilde{\gamma }}^{-1}_{obs\mathrm{|}MLMI,WB}$ approximates a scaled inverse chi-square variable with $df={\nu }_1$, but this inverse chi-square has the same CV as an ordinary chi-square variable with $df={\nu }_1-4$. So we can approximate ${\widetilde{\gamma }}^{-1}_{obs\mathrm{|}MLMI,WB}$ as a scaled ${\chi }^2_{{\nu }_1-4}$ variable. It follows that ${\widetilde{V}}_{MLMI,WB}$ is approximately scaled ${\chi }^2_{{\nu }_1-4}$ as well.

Now
\begin{equation}
\widetilde{V}_{MLMI,WB}={\widetilde{V}}_{ML|MLMI,WB}+\frac{1}{M}{\widehat{B}}_{MLMI}
\end{equation}

\noindent is the sum of two scaled chi-square variables with respective $df$s equal to ${\nu }_1-4$ and $M-1$. The variables are not independent, but the covariance between them is negligible if $M$ is large or ${\gamma }_{mis}$ is small. If we apply the Satterthwaite approximation to the sum, we get expression \eqref{df_MLMI_WB} for the $df$ of ${\widetilde{V}}_{MLMI,WB}$.
 
\section{Wang \& Robins' SB estimators}
\label{appendix:SB_estimator_WR}

In section \ref{section:SB_variance_estimation} we mentioned that Wang and Robins \cite{Wang1998}, Lemma 2, use a different SB estimator for $V^{-1}_{ML}$. After correction of a typo,\footnote{ Wang and Robins inadvertently divide $V^{-1}_{ML|SB}$ by $N$.} their estimator is 
\begin{equation}
\check{V}^{-1}_{ML|SB} = \frac1{M(M-1)} \sum_{m \neq m'} \sum^N_{i=1} c_{mm',i}
\end{equation}

\noindent where

\begin{equation}
c_{mm',i} = \frac1{2}({{\widehat{s}}^T_{com,m,i}\widehat{s}}_{com,m',i}+{{\widehat{s}}^T_{com,m',i}\widehat{s}}_{com,m,i}) 
\end{equation}

\noindent is the ``symmetrized'' cross-product of score estimates between one SI dataset ($m$) and another ($m'$). The cross-product ${{\hat{s}}^T_{com,m,i}\hat{s}}_{com,m',i}$ is not symmetric, and neither is the reverse cross-product ${{\hat{s}}^T_{com,m',i}\hat{s}}_{com,m,i}$, but the average $c_{mm'}$ is symmetric and so can be used to estimate the symmetric matrix $V^{-1}_{ML}$. 

Since $c_{mm'}=c_{m'm}$ we can halve the number of cross-products we need to calculate by restricting ourselves to cross-products where $m<m'$. Then Wang and Robins' estimator simplifies to
\begin{equation} 
\check{V}^{-1}_{ML|SB} = \frac{2}{M(M-1)}\sum_{m<m'}{\sum^N_{i=1}{c_{mm',i}}}
\end{equation}

${\check{V}}^{-1}_{ML|SB}$ looks quite different from our estimator ${\hat{V}}^{-1}_{ML|SB}$, but in fact the two are just different formulas for estimating the between-group variance of ${\hat{s}}_{com,m,i}$. To see this, notice that, if ${\hat{s}}_{com,m,i}$ is scalar, then ${\check{V}}^{-1}_{ML|SB}$ becomes
\begin{equation}
{\check{V}}^{-1}_{ML|SB} = \frac{2}{M(M-1)}\sum_{m<m'}{\sum^N_{i=1}{{{\hat{s}}_{com,m,i}\hat{s}}_{com,m',i}\ }}
\end{equation}

\noindent which, if divided by $N$ and $V({\hat{s}}_{com,m,i})$, is just a century-old formula for estimating the intraclass correlation \cite{Harris1913,Fisher1925}.\footnote{The old formula would center ${\hat{s}}_{com,m,i}$ around its sample mean, but that is not necessary here since we know that the mean of ${\hat{s}}_{com,m,i}$ is zero.} The intraclass correlation formula can be simplified so that no cross-products are required \cite{Harris1913}; applying the simplification, we get 

\begin{equation}
\check{V}^{-1}_{ML|SB} = \frac{M}{M\mathrm{-1}}\sum_{m<m'}{\sum^N_{i=1}{{\left({\overline{s}}_{com,i}\right)}^{\otimes 2}-\frac{1}{M-1}\ {\hat{V}}^{-1}_{com|SB}}}
\end{equation}

\noindent which is very similar to our ${\hat{V}}^{-1}_{ML|SB}$.

\end{document}